\documentclass[12pt,tightenlines,eqsecnum,floats,showpacs,nofootinbib,aps,prd]{revtex4}

\usepackage[colorlinks,citecolor=blue,linkcolor=black,urlcolor=blue]{hyperref}
\usepackage{enumitem}
\usepackage{graphicx,verbatim}
\usepackage{amsmath}
\usepackage{amsfonts}
\usepackage{amssymb}
\usepackage{epstopdf}
\usepackage[strings]{underscore}

\def\t{\tilde}
\def\h{\hat}
\def\b{\bar}

\def\ul{\underline}
\def\={\hat{=}}
\def\f{\frac}
\def\Im{{\rm Im}}
\def\Re{{\rm Re}}
\def\rmd{\mathrm{d}}
\def\l{\ell}

\def\Lie{\mathcal{L}}

\def\scri{\mathfrak{I}}
\def\scrip{\scri^{+}}
\def\scrim{\scri^{-}}

\def\4z{\Psi_{4}^{\circ}}
\def\3z{\Psi_{3}^{\circ}}
\def\2z{\Psi_{2}^{\circ}}
\def\1z{\Psi_{1}^{\circ}}
\def\0z{\Psi_{0}^{\circ}}
\def\sigmaz{\sigma^{\circ}}
\def\sigmabz{\bar{\sigma}^{\circ}}
\def\iz{i^{\circ}}

\def\qo{\mathring{q}}

\def\Do{\mathring{D}}
\def\no{\mathring{n}}

\def\lo{\mathring{\ell}}
\def\mo{\mathring{m}}
\def\mbo{\mathring{\b{m}}}
\def\d2vo{\rmd^{2} \mathring{V}}
\def\too{\mathring{t}}
\def\Fo{\mathring{F}}

\newcommand{\pb}[1]{\hbox{\lower0.5ex\hbox{${}_{\leftarrow}$}}\kern-1.9ex{#1}}

\def\S{\mathcal{S}}

\def\Lor{\mathfrak{L}}

\def\B{\mathfrak{B}} 
\def\P{{\mathfrak{p}}}

\def\T{{\mathcal{T}}}
\def\F{{\mathcal{F}}}
\def\iz{i^{\circ}}
\def\J{\vec{J}}
\def\Ji{\vec{J}_{\iz}}
\def\Sf{\vec{S}_{i^{+}}}
\def\Poi{{\P}{(\iz)}}
\def\Pof{\P{(i^{+})}}
\def\momi{P_{(\alpha)}{(\iz)}}

\def\s{s}

\def\be{\begin{equation}}
\def\ee{\end{equation}}
\def\ba{\begin{eqnarray}}
\def\ea{\end{eqnarray}}

\begin{document}

\title{Compact binary coalescences: \\The subtle issue of  angular momentum }
 
 \author{Abhay Ashtekar}
\email{ashtekar.gravity@gmail.com} 
\author{Tommaso De Lorenzo}
\email{tdelorenzo@psu.edu} 
\author{Neev Khera}
\email{neevkhera@psu.edu}
\affiliation{Institute for Gravitation and the
Cosmos \& Physics Department, Penn State, University Park, PA 16802,
U.S.A.}

\begin{abstract}

In presence of gravitational radiation, the notion of angular momentum of an isolated system acquires an infinite dimensional supertranslation ambiguity. This fact has been emphasized in the mathematical general relativity literature over several decades.  We analyze the issue in the restricted context of compact binary coalescence (CBC) where the initial total angular momentum of the binary and the final black hole spin generically refer to \emph{distinct} rotation subgroups of the Bondi-Metzner-Sachs group, related by \emph{supertranslations}.  We show that this ambiguity can be quantified using gravitational memory and the `black hole kick'. Our results imply that, although the ambiguity is conceptually important,  under assumptions normally made in the CBC literature, it can be ignored in practice for the current and foreseeable gravitational wave detectors.

\end{abstract}

\maketitle
\section{Introduction}
\label{s1}
The goal of this paper is to resolve a conceptual tension in the literature on angular momentum of isolated gravitating systems, in the context of Compact Binary Coalescences (CBCs).  

In presence of gravitational waves, ripples in space-time curvature persist all the way to null infinity, $\scrip$, and introduce an ambiguity in the notion of rotations and boosts even in the asymptotic region. Thus, even though space-time is asymptotically Minkowskian, the asymptotic symmetry group at $\scrip$ is not the Poincar\'e group $\P$, but an infinite-dimensional generalization thereof, the Bondi-Metzner-Sachs (BMS) group $\B$: While $\B$ is structurally similar to $\P$, the 4-dimensional subgroup $\T$ of  translations in $\P$ is replaced by an \emph{infinite} dimensional subgroup $\S$ of supertranslations of $\B$. Consequently, whereas $\P$ admits a 4-parameter family of Lorentz subgroups --related to one another by translations-- $\B$ admits an infinite parameter family of Lorentz subgroups, related to one another by supertranslations (see, e.g., \cite{sachs2,bondi-sachs,rg,aa-yau}).  Since angular momentum refers to  the Lorentz group, the relativistic angular momentum $M_{ab}$ in Minkowski space physics comes with a 4-parameter ambiguity which corresponds precisely to the choice of an origin about which angular momentum is defined. In asymptotically Minkowski space-times, by contrast, the ambiguity in angular momentum is \emph{infinite-dimensional} and cannot be traced to the choice of an origin in space-time.  This  dramatic shift occurs because of gravitational waves. In absence of gravitational waves, one can naturally reduce $\B$ to a Poincar\'e subgroup $\P$ thereof  \cite{np2,aa-rad}  and the supertranslation ambiguity disappears. Similarly, since gravitational waves do not reach spatial infinity, $\iz$, one can again reduce the asymptotic symmetry group at $\iz$ to the Poincar\'e group, and introduce the familiar notion of angular momentum there \cite{aarh,aa-io}. 
 
The situation at $\scrip$ came as a major surprise when it was first discovered and, in the subsequent decades, generated substantial literature  aimed at introducing a conceptually meaningful notion of angular momentum at $\scrip$ (see, e.g., \cite{jw,bb1,crp,ms-thesis,ms,jw-rev,aams,rgjw,aajw,rp2,shaw,ds,td,wz,ak-thesis,aaak}).  {The challenge was two-fold.}   On the mathematical side, the task was to  find expressions of angular momentum of the system at a retarded instant of time --represented by a cross-section of $\scrip$-- and of the flux of angular momentum carried by gravitational waves across any sub-region $\Delta\scrip$ of $\scrip$. On the conceptual side, the issue was whether the supertranslation ambiguity is avoidable.  As for concrete expressions of angular momentum and its flux, initially there was considerable confusion and many of the early expressions had  unphysical features. In particular,  in most cases the flux of BMS angular momentum through a patch $\Delta \scrip$ bounded by two generic cross sections was non-zero \emph{in Minkowski space} \cite{jw,bb1,crp,ms-thesis,ms,jw-rev,rgjw}!%
\footnote{in general, the past and future cross-sections of $\scrip$ that are used to define the initial angular momentum  $\Ji$ of the binary and the final spin $\Sf$ of the black hole are related by a supertranslation rather than a time-translation. This is why, as a check,  it is important to allow generic cross-sections in Minkowski space which are also related by a general supertranslation. If a flux formula yields a non-zero flux in Minkowski space for such cross-section, it is difficult to have faith in the flux it yields in generic situations of physical interest.}
The situation was subsequently clarified and a satisfactory expressions of the Bondi angular momentum and its flux  are since then available \cite{aams,td,wz,ak-thesis,aaak}. They have all the necessary mathematical invariances as well as  expected physical properties (summarized in the last Section of  Ref. \cite{aajw}). On the conceptual side, by now it is widely recognized in the mathematical General Relativity (GR)  community  that  the underlying supertranslation ambiguity cannot be avoided in presence of gravitational waves:  one just has to live with the  `infinite-dimensional' \emph{BMS angular momentum}.

However, the supertranslation ambiguity is generally ignored  in the CBC community (see, e.g., \cite{blanchet1,lbgf,lz1,lz2,lz3,nblr,tdanetal} {for examples of discussion of angular momentum}).  In particular,  the fact that the initial total angular momentum $\Ji$ of the binary and the spin $\Sf$ of the final black hole generically refer to \emph{different} ${\rm SO(3)}$ subgroups of $\B$, \emph{related by a supertranslation,}  is not  taken into account. If one restricts oneself to a ${\rm SO(3)}$ subgroup of $\B$, say the one adapted to the distant past, one can indeed introduce 3-vectors representing the angular momentum of the system at  retarded instants of time, and the flux of this angular momentum  carried by gravitational waves, both regarded as 3-vectors in an asymptotic Minkowski space.  But, since  the past  ${\rm SO(3)}$ is generically related to the future one by a  supertranslation, one would not obtain the correct black hole spin $\Sf$ in the distant future using a simple balance law that does not take into account the supermomentum carried by gravitational waves. Since supermomenta do not enter the angular momentum considerations of the CBC community, there is a clear conceptual tension.

The tension has persisted over the years, primarily because the supertranslation ambiguity  has not been \emph{quantified}, whence its observational  significance  has remained obscure.  The purpose of this paper is to change this status-quo by quantifying the ambiguity in the context of compact binaries emitting gravitational waves. We will present a systematic procedure to calculate the supermomentum that must be taken into account in angular momentum considerations of exact GR. The result will  show explicitly that, although the ambiguity is conceptually important, we have the happy circumstance that one can ignore it \emph{in practice.} More precisely, because of the asymptotic boundary conditions that are normally imposed at $i^{\circ}$ and $i^{+}$ in the analysis of CBCs, the supermomentum contribution is  small for the kick velocities normally considered, orders of magnitude smaller than the statistical errors associated with detectors. 

We have made a special effort to address both the waveform and the mathematical GR communities in order to bring the discussion to a common platform. In Section \ref{s2} we recall some results that will provide the conceptual basis for the rest of the paper.  In particular, we summarize the relation between rest frames, rotation subgroups of the Poincar\'e and BMS groups, and the commonly used  angular momentum 3-vectors.  { (Mathematical relativists can skip this discussion.)}  In Section \ref{s3} we discuss the BMS angular momentum at $\scrip$ for CBC.   We will find that the asymptotic conditions in distant past and distant future enable us to single out \emph{two} Poincar\'e subgroups, $\Poi$ and $\Pof$ of the BMS group: The initial angular momentum $\Ji$ refers to $\Poi$ while the final spin $\Sf$ refers to $\Pof$. The two Poincar\'e groups are distinct unless the (total) gravitational memory vanishes,  and {are} related by a supertranslation.  While comparing $\Ji$ with $\Sf$, one has to take into account this supertranslation as well as the fact that the past and future rest frames are in general different because of the black hole recoil or kick \cite{kicks1,kicks2}. We will show that this extra term  can be computed directly from the waveform.  In Section \ref{s4} we summarize the main result and discuss why the supermomentum contribution turns out to be negligibly small under assumptions normally made by the CBC community.  In Appendix A we show that all results of Section \ref{s3}  -- expressions of $\Ji$, $\Sf$ and fluxes relating them-- continue to hold under weaker assumptions on the behavior of the system in the distant past and distant future. However, now we can no longer conclude that the supermomentum ambiguity is negligible. Therefore, if it should turn out that the weaker asymptotic conditions at $\iz$ and $i^{+}$ are needed in the analysis of CBCs of physical interest, the issue of importance of the supermomentum term will have to be revisited.

We use the same notation as in the companion paper \cite{adlk1} but set $c=1$. For convenience of readers who are interested only in the issue of angular momentum --rather than the measures of accuracy of waveforms discussed in \cite{adlk1}--   we have attempted to make this paper essentially self-contained.  {As in \cite{adlk1}, the term `mathematical relativity' is used to refer to the literature (such as Refs.  \cite{jw,bb1,crp,ms-thesis,ms,jw-rev,aams,rgjw,aajw,rp2,shaw,ds,td,wz,ak-thesis,aaak}) that assumes that space-times under consideration admit a conformal completion at $\scri$ a la Penrose  \cite{rp}  and works out the consequences. Indeed, this degree of regularity  is needed for the notion of angular momentum to be well-defined at null infinity in the first place. We will briefly discuss the issue of existence  of such space-times in Section \ref{s4}.} 

\section{The supertranslation ambiguity} 
\label{s2}

This section is divided in three parts. In the first we fix terminology and recall the notion of relativistic angular momentum. In the second, we pinpoint the difficulty encountered in extending this notion to $\scrip$. In the third, we recall how this obstacle can be overcome in \emph{absence  of} gravitational waves (in particular in stationary space-times).  In Section \ref{s3}, we will apply these ideas to CBCs {under the standard assumption made in the CBC community that the system becomes} asymptotically stationary (in a certain weak sense, specified in Section \ref{s3.1}).

\subsection{Relativistic angular momentum}
\label{s2.1}

Let us begin by recalling the notion of angular momentum in special relativity. Let $(M, \eta_{ab})$ be Minkowski space-time. A constant, unit time-like vector field $\tau^{a}$ is said to fix a \emph{Lorentz frame} because it represents the 4-velocity of a family of inertial observers.  There is a  three parameter family of these observers, related to one another by Lorentz boosts which map $\tau^{a}$ to another constant, unit time-like vector field $\tau^{\prime\,a}$.  Next, recall that by fixing an origin $O$, the 10 Killing fields $K^{a}$ of the Minkowski metric can be written as 
\be \label{killing} K^{a} \,=\, \too^{a} + \Fo^{ab}X_{b} \, \equiv\, \too^{a} + L^{a} \ee
where $\too^{a}$ is a constant vector field --a translation Killing field--  and $L^{a}$ a Lorentz Killing field constructed from a constant skew symmetric tensor field $\Fo^{ab}$  and  the position vector $X^{a}$ of the point at which $K^{a}$ is evaluated, relative to $O$. Thus, while we have a well-defined notion of a `pure' translation $\too^{a}$,  we can speak of a `pure' Lorentz transformation $L^{a}$ only relative to an origin. Now, given a physical system with conserved stress-energy tensor $T_{ab}$, the 10 Poincar\'e generators $K^{a}$ enable us to define 10 conserved quantities, namely $P_{a}$ the 4-momentum, and $M_{ab}$ the relativistic angular momentum:
\be P_{a}\too^{a} + M_{ab} \Fo^{ab} \,:=\, \int_{\Sigma} T_{ab} K^{b} \, \rmd S^{a}  \, ,\ee
 where the integral is performed on a Cauchy surface $\Sigma$ of Minkowski space. We will restrict ourselves to the physically interesting case where $P_{a}$ is time-like; $P_{a} = -M_{o} \tau_{a}$, where $M_{o}>0$ is the rest mass: $M_{o}^{2} = - P_{a}P^{a}$. We will refer to the Lorentz frame defined by this $\tau^{a}$ as \emph{the rest frame of the system}.

 Under a displacement  $O \to O^{\prime}$ of the origin we have
 \be \label{trans} P_{a}  \to P_{a} \quad {\rm and} \quad M_{ab} \to M_{ab}\, + \,M_{o}\, \tau_{[a}  d_{b]}  \ee
where $d^{a}$ is the position vector of $O^{\prime}$ relative to $O$.   Thus,  three of the six components of $M_{ab}$ --corresponding to the three boosts in the rest frame of the system-- can be transformed away by change of origin. The non-trivial information corresponds just to the  \emph{rotation subgroups of the Lorentz group selected by the rest frame} of the system. This is encoded in the angular-momentum spatial vector $\J^{d}$:
\be \label{J1}  \J^{d} := \epsilon^{abcd} \,\tau_{a}\, M_{bc} \ee  
Thus, although we have 10 conserved quantities, physical information they carry is encoded entirely in the 4-momentum $P_{a}$ and the angular momentum 3-vector $\J^{a}$.

\subsection{ The conceptual obstacle at $\scrip$}
\label{s2.2}

With these preliminaries out of the way, let us turn to null infinity of asymptotically Minkowski space-times (for notation and an introduction, see \cite{adlk1}; for precise definition, see e.g. \cite{aa-yau}). The key question for us is: can we define the analog of the 4-momentum $P_{a}$ and the angular momentum vector $\J^{a}$ at $\scrip$? Because gravitational waves carry energy-momentum and angular momentum, these notions can at best be time-dependent. Recall that each cross-section $C$ of $\scrip$ represents a retarded instant of time. So the question is whether we can meaningfully associate a 4-momentum and  angular momentum to each cross-section $C$ of $\scrip$ in a consistent manner.

Let us begin with the kinematical structures at $\scrip$ that have direct analogs in the geometry of Minkowski space. First, $\scrip$ is equipped with a 3-parameter family of pairs $(\qo_{ab}, \no^{a})$ of fields, where $\qo_{ab}$ is the unit 2-sphere metric, and $\no^{a}$ a null normal  to $\scrip$  --the limit to $\scrip$ of an \emph{asymptotic} (unit) time-translation Killing field $\tau^{a}$ in the physical space-time. Each such pair is referred to as a \emph{Bondi (conformal) frame}. Thus, there is a 1-1 correspondence between the \emph{asymptotic} Lorentz frame in the physical space-time selected by $\tau^{a}$, and the Bondi frame $(\qo_{ab}, \no^{a}$) on $\scrip$.  If we first fix a Lorentz frame and then perform a boost corresponding to a velocity $\vec{v}$,  then $\tau^{a}$ is mapped to another unit asymptotic time translation $\tau^{\prime\, a}$ and the initial Bondi-frame $(\qo_{ab}, \no^{a})$ transforms as 
\be (\qo_{ab}, \, \no^{a}) \to (\qo_{ab}^{\prime}, \, \no^{\prime\,a}) \,=\, (\omega^{2} \qo_{ab}, \, \omega^{-1}\no^{a}), \quad {\rm with} \quad \omega = \f{1}{\gamma(1 - \vec{v}\cdot \h{x})} \ee
where $\gamma = (1 -v^{2})^{-\f{1}{2}}$ is the standard Lorentz factor, and $\hat{x}^{a}$ is the unit radial vector with Cartesian components $(\sin\theta\cos\varphi, \, \sin\theta \sin\varphi, \, \cos\theta)$ in the $(\theta,\varphi)$-chart adapted to $\qo_{ab}$. (Recall we have set $c=1$.) 

Next, the role of Killing fields $K^{a}$ in Minkowski space is assumed by the generators $\xi^{a}$ of the BMS group.  In any given Bondi-frame $(\qo_{ab},\, \no^{a})$ the explicit form of these vector fields on $\scrip$ is given by \cite{sachs2,np2,ak-thesis}:
\be \label{generators} \xi^{a}(u,\theta,\varphi) \, =\, \big[f(\theta,\phi) + u\, k(\theta,\phi) \big]\no^{a} \,+\, \qo^{ab}\Do _{b}k(\theta,\phi) + {\mathring{\epsilon}}^{ab} \Do_{b} \beta(\theta,\phi)\, , \ee
where $f$ is any smooth function on a 2-sphere, $u$ is an affine parameter of $\no^{a}$ (i.e., $\no^{a}\partial_a u =1$), $\qo^{ab}$ and ${\mathring\epsilon}^{ab}$ are the metric and the alternating tensor on the $u=const$ cross-sections of $\scrip$, and $k(\theta,\varphi)$ and $\beta(\theta,\varphi)$ are linear combinations of the $\ell=1$ spherical harmonics (defined by the unit, round metric $\qo_{ab}$). The first term, $f \no^{a}$, of $\xi^{a}$ represents a BMS supertranslation, which reduces to a BMS translation if $f(\theta,\varphi)$ is a linear combination of the first four spherical harmonics.  We will denote the BMS translations by $\alpha\no^{a}$, to distinguish them from generic supertranslations $f\no^{a}$. Although we have used a Bondi-frame in this description, the resulting supertranslation and translation subgroups  $\S$ and $\T$  are independent of this choice. 

The interpretation of the rest of $\xi^{a}$, on the other hand, makes use of the specific Bondi-frame and choice of $u$. Recall first that the Bondi-frame corresponds to an asymptotic rest frame. The 1-parameter family of cross-sections $u=const$ is the analog of a world-line of the  time-translation Killing field $\tau^{a}$ in Minkowski space that defines the rest frame under consideration. Now, for a fixed $\no^{a}$,  there is a (supertranslation) freedom $u\to \t{u} = u + \s (\theta,\phi)$ in the choice of $u$, where $\s$ is any smooth function. Therefore, while given a rest frame $\tau^{a}$ in Minkowski space we have only a three-parameter family integral curves of $\tau^{a}$, each representing a `candidate' center of mass world lines of the given system, given a rest frame $(\qo_{ab}, \no^{a})$ at $\scrip$ we have an infinite parameter family of `candidate' center of mass `world-lines', each represented by the family $u=const$ of cross-sections that are preserved by $\no^{a}$.  The three vector fields ${\mathring{\epsilon}}^{ab}   \Do_{b} \beta(\theta,\phi) $ define `the'  rotation or ${\rm SO(3)}$ subgroup in the rest and center of mass frame determined by $\no^{a}$ and the choice of $u$. The remaining three,  $u k(\theta,\phi) \no^{a} \,+\, \qo^{ab} \Do_{b}k(\theta,\phi) $ represent `the' boosts in this frame. 

Note that the last six vector fields --which together generate a Lorentz subgroup  $\Lor$ of $\B$-- are tangential to precisely one cross-section, namely, $u=0$. Now,  given \emph{any} cross-section $C$ of $\scrip$ we can adapt the affine parameter $u$ of $\no^{a}$ to it so that  $u=0$ on that $C$. Therefore each cross-section $C$ of $\scrip$ picks out a Lorentz subgroup $\Lor_{C}$ of $\B$. Furthermore, this is a 1-to-1 correspondence.  Finally, since any cross-section $C$ can be mapped to any other cross-section $C^{\prime}$ by a supertranslation, any two Lorentz subgroups are also mapped to one another by a supertranslation. In this precise sense,  $\B$ admits `as many' Lorentz subgroups as there are supertranslations. By contrast, in Minkowski space-time, there is a natural 1-1 correspondence between space-time points and Lorentz subgroups $\Lor$ of the Poincar\'e group $\P$ --the action of each subgroup leaving precisely one point invariant--  whence $\P$ admits `as many' Lorentz subgroups as there are translations.

{Now, associated with each BMS vector field $\xi^{a}$ there is a quantity $P_{\xi}[C]$ representing the $\xi$-component of the Bondi-momentum at the retarded instant of time defined by the cross-section $C$ of $\scrip$. The explicit expressions of $P_{\xi}[C]$ involve certain physical fields at $\scrip$, in addition to the symmetry vector field $\xi^{a}$. These fields  are the asymptotic shear $\sigmaz$,
\be\label{shear}
\sigmaz(u,\theta,\varphi)  = \f{1}{2} \lim_{r\to \infty} \, r\, \big(h_{+} +i h_{\times}\big) (u,\theta,\varphi), \ee
constructed from the wave form, and the Newman-Penrose fields $\2z$ and $\1z$, 
\be \label{psi} 2\2z (u,\theta,\varphi) = \displaystyle{\lim_{r\to \infty}\,  r^{3} C_{abcd} (n^{a} {\l}^{b} n^{c}  {\l}^{d} + n^{a} {\l}^{b} m^{c}  {\b{m}}^{d}}); \qquad  \1z (u,\theta,\varphi) = \displaystyle{\lim_{r\to \infty}\,  r^{4} C_{abcd} \l^{a} {m}^{b} \l^{c}  {n}^{d}}\, \ee
constructed from the asymptotic Weyl curvature. Here, $n^{a}, \l^{a}, m^{a}, \b{m}^{a}$ is a Newman-Penrose tetrad adapted to the Bondi-frame and the cross section $C$  (given by $u=u_{1}$ in Fig. 1),  and the limit is taken along outgoing null cones  ($u={\rm const}$).  See, e.g., \cite{adlk1} for further discussion.}

With this structure at hand, we can introduce the 4-momentum and supermomentum. 
The 4-momentum $P_{(\alpha)}[C]$ is the Bondi-momentum $P_{\xi}[C]$ associated with the BMS  translation $\xi^{a} = \alpha(\theta,\varphi) \no^{a}$ at $\scrip$  \cite{sachs1,bondi-sachs,np} :
\be \label{mom} P_{(\alpha)} [C]   := - \f{1}{4\pi G} \,\oint_{C}\!\d2vo\, \alpha(\theta,\varphi)\,\, \Re \big[\2z +\sigmabz \dot{\sigma}^{\circ} \big](\theta,\varphi)\, , \ee
where $\d2vo$ is the volume element of a  unit 2-sphere and $\no^{a}$ is the limit of $n^{a}$ to $\scrip$. Now, because the BMS translations constitute a canonical 4-dimensional (normal) subgroup of the BMS group \cite{sachs2}, we know what it means to consider the same BMS translation on another cross-section $C^{\prime}$. The energy-momentum flux  $\F_{(\alpha)}$ carried by gravitational waves across the region $\Delta\scrip$ bounded by these cross-sections is given by the difference between the two Bondi 4-momenta \cite{sachs1,bondi-sachs,np}
\be
\begin{split}  \label{flux1}  
P_{(\alpha)} [C^{\prime}] - P_{(\alpha)} [C]   \equiv   \F_{(\alpha)}  :&= \, -\, 
\f{1}{8\pi G} \int_{\Delta\scrip}\!\!\!\! \rmd u\,\, \d2vo\,\, \dot{\sigma}^{\circ\, ab} \big(\Lie_{\alpha\no} \,\sigma^{\circ}_{ab}  + \Do_{a}\Do_{b}f  \big)\\
&=-\, \f{1}{4\pi G} \int_{\Delta\scrip}\!\!\!\! \rmd u\,\, \d2vo\,\, \alpha(\theta,\varphi) \, |\dot\sigma^{\circ}|^{2}(u,\theta,\varphi)\, ,
\end{split}
\ee
if $C^{\prime}$ is to the future of $C$, where $\sigma^{\circ}_{ab} = - (\b{\sigma}^{\circ} m_{a}m_{b} + \sigma^{\circ}\b{m}_{a}\b{m}_{b})$ and $\Do$ the derivative operator compatible with $\qo_{ab}$. Finally, in view of the positive energy theorem at null infinity \cite{sy,hp,rt}, it is customary to assume that the Bondi 4-momentum $P_{(\alpha)} [C] $ is a time-like vector. It provides us with \emph{an instantaneous rest frame} of the system, corresponding to that time instant. Thus the situation with 4-momentum at $\scrip$ is completely analogous to that in special relativity, except that the Bondi 4-momentum is not conserved; it refers to a retarded instant of time. The balance law (\ref{flux1}) is often used in CBC --for example, the final black hole kick is estimated by evaluating the flux of the 3-momentum across $\scrip$ (in the initial rest frame) using the right hand side of (\ref{flux1}) \cite{kicks1,kicks2}. 

\begin{figure}[t]
\center
\includegraphics[width=.7\textwidth]{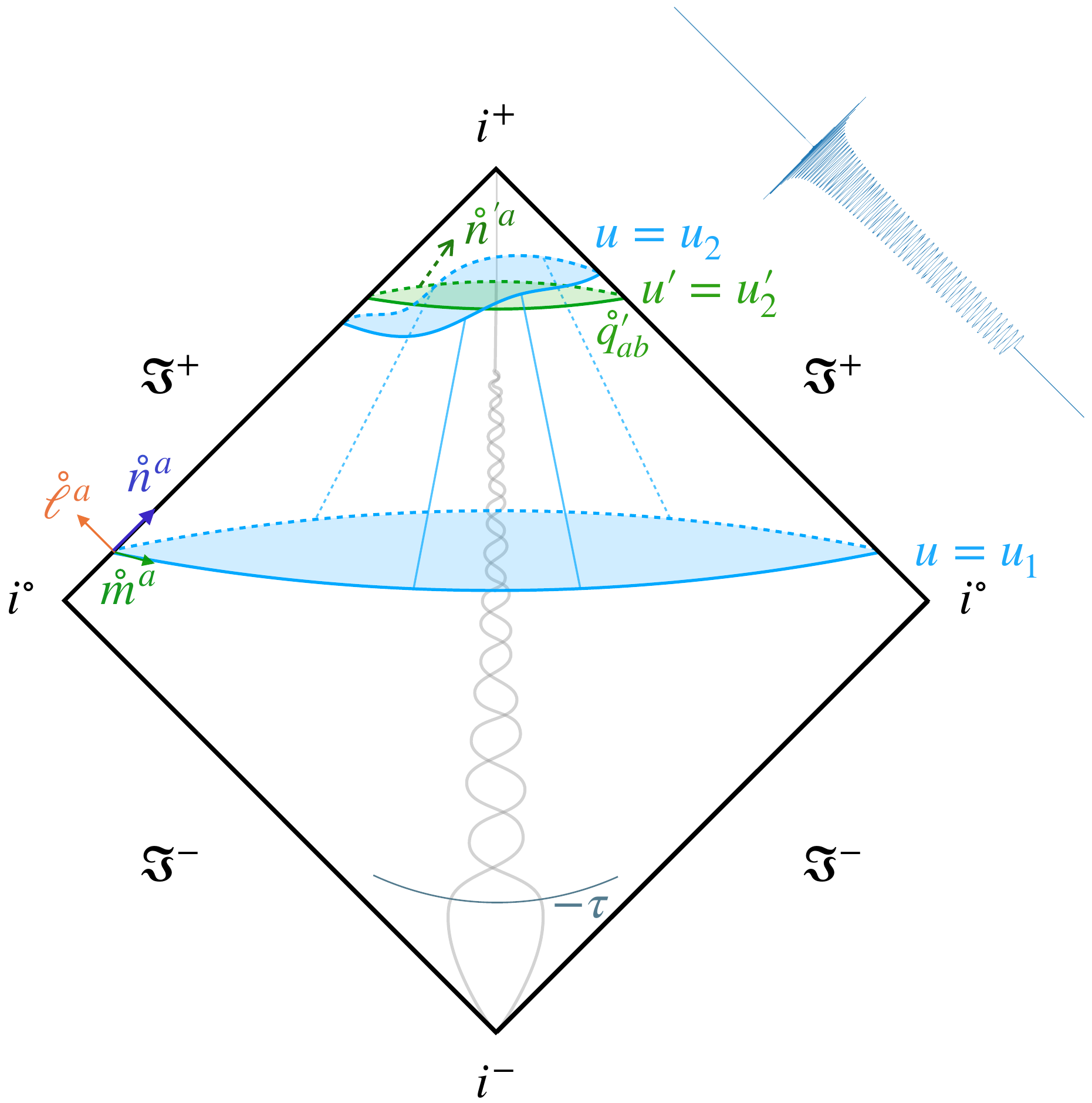} 
\caption{A depiction of a compact binary space-time, together with an artist's impression of the wave form based on \cite{bohe2017improved}.  $\scrip$ constitutes the future boundary of this space-time and has topology $\mathbb{S}^{2}\times \mathbb{R}$. The Bondi news --the time derivative of the waveform-- goes to zero in the distant past and distant future.  Because of the black hole kick, the Bondi (conformal) frames $(\qo_{ab}, \no^{a})$ and $(\qo_{ab}^{\,\prime}, \no^{\prime\, a})$ adapted to rest-frames in the distant past and distant future are distinct. Cross-sections $u=u_{1}$ and $u=u_{2}$ belong to the center of mass family adapted to the distant past; the shear of this family of cross-sections vanishes as $u\to -\infty$. They define the past Poincar\'e group $\Poi$ discussed in Section \ref{s3.1}. However, generically their shear fails to vanish in the distant future. There is a distinct family of cross-sections --such as $u^{\prime} = u_{2}^{\prime}$--  that are adapted to the rest frame in the future and become shear-free in the limit $u\to \infty$.  This family defines the future Poincar\'e group $\Pof$. Generically the two Poincar\'e groups are distinct subgroups of the BMS group $\B$, related by a BMS \emph{supertranslation}. The total angular momentum $\Ji$ in the distant past refers to $\Poi$, while the final spin $\sf$ refers to $\Pof$. Therefore, the difference $\Ji - \Sf$ involves flux of angular momentum \emph{as well as} supermomentum.}
 \label{Fig1}
\end{figure}

Finally, associated with any supertranslation $\xi^{a} = f(\theta,\phi) \no^{a}$ there is a supermomentum $P_{(f)}[C]$ for any cross-section $C$ and a flux $\F_{(f)}$ associated with any region $\Delta\scrip$:
\ba
 \label{supmom}
\begin{split} P_{(f)} [C]   :=& - \f{1}{4\pi G} \,\oint_{C}\!\d2vo\,  f(\theta,\varphi)\, \Re \big[\2z + \sigmabz \dot\sigma^{\circ}\big](\theta,\varphi)\,\, , {\rm and}, 
\end{split}\\
\label{flux2}
\begin{split}   \F_{(f)} :=& \, -\,  \f{1}{8\pi G} \int_{\Delta\scrip}\!\!\!\! \rmd u\,\, \d2vo\,\, \dot{\sigma}^{\circ\, ab} \big(\Lie_{f\no} \,\sigma^{\circ}_{ab}  + \Do_{a}\Do_{b}f  \big)\\  
=& - \, \f{1}{4\pi G} \int_{\Delta\scrip}\!\!\!\! \rmd u\, \d2vo\, f(\theta,\varphi) \,
\big[ |\dot\sigma^{\circ}|^{2}  -  \Re(\eth^{2} \dot{\bar\sigma}^{\circ}) \big]  (u,\theta,\varphi)\, .\end{split}
\ea
Here $\eth$ is the usual angular derivative: If $A$ has spin weight $s$, then $\eth A$ is a field with spin-weight $s+1$ given by
\be
     \eth A =  \f{1}{\sqrt{2}}\, (\sin\theta)^{s}\,  \big(\partial_{\theta} + \f{i}{\sin\theta}\, {\partial_{\varphi}} \big)\, (\sin\theta)^{-s}\, A\, .
\ee

Let us now turn to angular momentum. One immediately encounters an obstacle: Since the BMS 
group $\B$ admits an infinite-parameter family of Poincar\'e subgroups $\P$ rather than one, we cannot just repeat the familiar procedure from special relativity outlined above to obtain an analog of $\J^{a}$. However, given any cross-section $C$ we do have a preferred Lorentz subgroup $\Lor$ of $\B$ which we can use to construct the analog of the relativistic angular momentum tensor $M_{ab}$. Furthermore, the Bondi 4-momentum $P_{(\alpha)}[C]$ provides the instantaneous rest frame for the system at $C$. Can we not put these two elements together to arrive at the desired analog $\J^{a}[C]$ of $\J^{a}$?

Indeed, this idea can be implemented in detail. The problem is that as we change the cross-section $C$ to another one, $C^{\prime}$,  \emph{the Lorentz group $\Lor_{C}$ as well as the rest frame determined by the 4-momentum $P_{(\alpha)}[C]$ changes}.  In the special case when $C$ is mapped to $C^{\prime}$ by a BMS translation, $\Lor_{C}$ and $\Lor_{C^{\prime}}$ are related by a translation as in special relativity. But even in this case the Bondi 4-momentum changes, whence the two \emph{rest frames are different}. Therefore, $\J^{a}$ at the two cross-sections would be associated with  \emph{different} ${\rm SO(3)}$ subgroups of $\B$.  Consequently, comparing $\J^{a}[C]$ with $\J^{a}[C^{\prime}]$ would be like comparing, in Minkowski space, the $\J_{z} \equiv M_{xy}$ component of angular momentum at a given time with, say, $M_{xy} + M_{zt}$ at another time! If we consider  two generic cross-sections, the situation becomes much worse. For, now the two cross-sections are related by a \emph{supertranslation}  (See Fig. 1), whence the analog of the transformation property (\ref{trans}) now involves the infinite component \emph{supermomentum} $P_{(f)}$ in place of the 4-momentum $P_{a}$.  Therefore, we are led to replace the 6-component object $M_{ab}$ in special relativity with an infinite component object  --the BMS angular momentum.  Generically, it is no longer possible to construct the analog of the 3-vector $\J^{a}$ of  Eq. (\ref{J1}) to encode  the full content of the BMS angular momentum. \emph{This is the celebrated `supertranslation ambiguity' in the notion of angular momentum.} 

As we remarked in the Introduction, in general we just have to live with it. But in classes of physical systems of interest --such as CBC-- one can significantly reduce the ambiguity. 

\subsection{Poincar\'e reduction of the BMS group}
\label{s2.3}
As a prelude to the discussion of CBCs in Section \ref{s3}, we will now summarize the procedure \cite{np2,aa-rad} that can be used to reduce the BMS group $\B$ to a canonical Poincar\'e subgroup $\P$ of $\B$ for systems that can be regarded as `stationary to leading order at $\scrip$'.  By themselves these systems are not of direct interest to CBCs because they do not admit any gravitational waves. Nonetheless, as we discuss in Section \ref{s3.1}, the main ideas can be generalized to situations in which the required condition is satisfied not on all of $\scrip$ but only asymptotically in time, i.e., in the limit $u\to \pm\infty$ on $\scrip$. This generalization is  directly applicable to CBCs.

Let us begin with stationary space-times $(M,g_{ab})$, and denote the time-translation Killing field (that has unit norm at infinity) by $\tau^{a}$. Then there is precisely one Bondi frame $(\qo_{ab}, \no^{a})$ at $\scrip$ whose $\no^{a}$ is the limit to $\scrip$ of $\tau^{a}$. Let us restrict ourselves to that Bondi frame. Then if $u$ denotes the affine parameter of $\no^{a}$, in the $(u,\theta,\varphi)$ chart on $\scrip$ we have $\no^{a} \partial_{a} = \partial/\partial u$ and $u$ can be taken to be the natural time variable at $\scrip$. In these space-times, the Bondi news $N = - \dot{\b\sigma}^{\circ}$ vanishes, signaling absence of gravitational radiation. This in turn implies that the components $\4z$ and $\3z$ of the asymptotic Weyl tensor vanish and $\dot{\Psi}_{2}^{\circ} =0$ at $\scrip$ (see, e.g., \cite{np,rg,aa-yau}). Next, since $\tau^{a}$ is a Killing field, in particular we have $\Lie_{\tau}  C_{abcd} =0$ everywhere on $M$. Therefore if we use a Newman-Penrose null tetrad $(\no^{a}, \lo^{a}, \mo^{a}, \mbo^{a})$ at $\scrip$ which is Lie dragged by $\no^{a}$, we  also have $\dot{\Psi}_{1}^{\circ}=0$, and  $\dot{\Psi}_{0}^{\circ} = 0$. 
\vskip0.1cm

The Poincar\'e reduction of $\B$ does not need stationarity, but a weaker, asymptotic version thereof. Let us consider space-times $(M,g_{ab})$ which are such that \vskip0.1cm
\noindent (I) The Bondi news vanishes, $N\equiv  -\dot{\bar\sigma}^{\circ}=0$ on $\scrip$; and, \\
(II) In the rest frame defined by the Bondi 4-momentum $P_{(\alpha)}$,\,  $\dot{\Psi}_{1}^{\circ} =0$ on $\scrip$.%
\footnote{Note that while Bondi news is invariant under the change of Bondi-frame, $\dot{\Psi}_{1}^{\circ}$ is not. Therefore condition (II) can be satisfied only in one Bondi frame. 
Now, since $N\equiv -\dot{\bar\sigma}^{\circ}=0$, Eq. (\ref{flux1})  implies that the Bondi 4-momentum is independent of the choice of cross-section.  Therefore there is a unique Bondi-frame in which the Bondi 3-momentum is zero; this is the rest frame of the system. Condition (II) is imposed in this Bondi-frame.}
\vskip0.1cm
We will refer to these gravitating systems as being  \emph{stationary to leading order at $\scrip$}.  Since vanishing of $N$ is conformally invariant, condition (I) implies that $\partial_{u}\sigmaz =0$ in \emph{any} Bondi frame $(\qo_{ab}, \no^{a})$ and for \emph{any} choice of the affine parameter $u$ of $\no^{a}$. Next, conditions (I) and (II) together with Bianchi identities imply that  $\Im \2z =0$ on $\scrip$  (and $\Re\2z$ is a constant  in the rest frame of the system) \cite{np,adlk1}.  Now, in any asymptotically Minkowskian space-time, $\Im \2z$ is completely determined by the asymptotic shear $\sigmaz$ and its derivatives:  $\Im \2z = 2\sqrt{2} \big(\Im \eth^{2} \b{\sigma}^{\circ} + \sigmaz  \dot{\bar\sigma}^{\circ} \big)$ \cite{np}. Vanishing of $\Im \2z$  and the Bondi news $N$ therefore implies
\be \label{electic}   \Im\, \eth^{2} {\bar\sigma}^{\circ} =0, \qquad {\hbox{\rm whose general solution is}} \quad \sigmaz = \eth^{2} \t{\s}    \ee
for some real, spin-weight zero function $\t{\s}$. Shears $\sigmaz$ of this form are said to be \emph{purely electric}. Now, using the fact that the Bondi news vanishes, one can also show that the shear $\sigmaz$ of two cross-sections $C$ and $C^{\prime}$ that are related to each other by a (finite) supertranslation $u \to  u  +  \s(\theta,\varphi)$ is given by
\be  \label{shearchange} \sigma^{\circ\, \prime} = \sigma^{\circ}  + \eth^{2} \s \, . \ee
Therefore, starting with \emph{any} cross-section one can make an appropriate supertranslation  to arrive at a shear-free cross-section:  in absence of Bondi news, \emph{purely electric shears can be transformed away by moving to a suitable cross-section}. Furthermore, the equation $\eth^{2} \s =0$ has precisely a 4-parameter family of solutions, each defining a BMS \emph{translation}. Therefore, it follows that $\scrip$ admits precisely a 4-parameter family of shear-free cross-sections --called \emph{good cuts}-- that are mapped into each other by BMS translations. Since each cross-section is left invariant by a Lorentz subgroup $\Lor$ of $\B$, the subgroup of  $\B$ that preserves this family of good cuts is a Poincar\'e group, say $\P_{o}$.  (Thus, as far as the Lorentz subgroups $\Lor$ of $\P_{o}$ are concerned, good cuts are the analogs of points in Minkowski space.)  Using these Lorentz subgroups, we can now define an angular momentum tensor $M_{ab}$ that transforms as in Eq. (\ref{trans}), where the `displacement vector' $d^{\,a}$ relating origins $O$ and $O^{\prime}$ is replaced by a (finite) BMS \emph{translation} that maps the first good cut $C$ to a second good cut $C^{\prime}$.

Next, as we remarked before, since $N=0$ in this case, the Bondi 4-momentum $P_{(\alpha)}[C]$ is independent of the choice of $C$ and we have a \emph{canonical} Bondi-frame at $\scri^{+}$ representing the asymptotic rest frame of the system. Thus both the obstacles encountered in Section \ref{s2.2} have been removed and one can define an angular momentum 3-vector $\J_{(\beta)}[C]$ for any cross-section $C$, which is again independent of the choice of $C$. (Here $\alpha$ is a linear combination of the first four spherical harmonics, representing a BMS translation and $\beta$ a linear combination of the $Y_{1,m}$ representing the direction of the axis of rotation. The explicit expression of $\J_{(\beta)}$ will be given in Section \ref{s3.2}.) Both $P_{(\alpha)}$ and $\J_{(\beta)}$ can be naturally regarded as co-vectors dual to the space of BMS translations; the first is time-like while the second is space-like and orthogonal to the first.

 To summarize, then,  the 4-momentum and angular momentum structure of  space-times that are stationary to leading order at $\scrip$  is the same as in special relativity.\\

\emph{Remarks:}
\begin{enumerate}
\item In Minkowski space-time $(M,\eta_{ab})$, each shear-free cross-section of $\scrip$  is the intersection of the future light cone of a point in the interior with $\scrip$. Thus, there is indeed a natural isomorphism between the 4-parameter family of points of Minkowski space and the 4-parameter family of good cuts on its $\scrip$. As one would expect, the Poincar\'e subgroup $\P_{o}$ is induced on $\scrip$ by the isometry group of Minkowski space. Finally, in this correspondence between the two Poincar\'e actions, the Lorentz subgroup $\Lor_{O}$ that leaves a point $O$ of Minkowski space invariant is sent to the  Lorentz subgroup $\Lor_{C_{O}}$ of $\P_{o}$ that leaves the good cut $C_{O}$ defined by $O$ invariant. In more general space-times that are only stationary to the leading order at $\scrip$, we have neither isometries in the interior, nor a natural correspondence between space-time points and shear-free cross sections.  But the shadows of some Minkowski structures cast on $\scrip$  --the Poincar\'e subgroup $\P_{o}$ of $\B$, the good cuts, and the 4-parameter family of Lorentz subgroups that leaves one of the good cuts invariant-- continue to exist.

\item We presented this procedure of Poincar\'e reduction using \emph{stationarity to leading order at $\scrip$} to bring out the physical motivation. From a mathematical perspective,  a necessary and sufficient condition for this procedure to go through is simply that the Bondi news $N$ and $\Im\2z$ must vanish on $\scrip$. These two conditions have an invariant geometric meaning. It turns out that  in any asymptotically Minkowskian space-time, $\scrip$ is naturally equipped with an equivalence class of intrinsically  defined connections $[D]$.%
\footnote{Without loss of generality one can assume that the conformal factor $\Omega$ in Penrose's completion is chosen such that the null normal $n^{a}$ to $\scrip$ is divergence-free. In such conformal frames, the space-time derivative operator compatible with the conformally rescaled metric can be pulled back to $\scrip$ to an intrinsically defined derivative operator $D$.  An equivalence class $[D]$ consists of various $D$ that are induced on $\scrip$ of any one physical space-time  through various choices of divergence-free conformal factors $\Omega$.}
The non-trivial part of its curvature is encoded precisely in $N$ and $\Im\2z$. Therefore the curvature is trivial (as on Minkowski $\scrip$) precisely when they vanish. Borrowing terminology from Yang-Mills theory, such connections are called \emph{classical vacua,} denoted by $[D_{\circ}]$ \cite{aa-rad,aa-bib}. There are as many vacua $[D_{\circ}]$ as there are supertranslations modulo translations. Each $[D_{\circ}]$ is left invariant under the action of a Poincar\'e subgroup of the BMS group. Thus, to single out a Poincar\'e subgroup, we need to select a specific classical vacuum. Each $[D_{\circ}]$ provides a 4-parameter family of good cuts of $\scrip$ and vice versa. But the invariant geometric meaning of various constructions in the next section are more transparent in terms of $[D_{\circ}]$. (For a summary, see \cite{aa-yau}.) In this paper we chose to emphasize `good cuts' in place of $[D_{\circ}]$ because researchers working with waveforms are likely to be more familiar with `good cuts'.
\end{enumerate}

\section{Compact binary Coalescence: Angular momentum at $\scrip$} 
\label{s3}

This section is divided into three parts. In the first we specify the class of systems we wish to consider: isolated gravitating bodies that become stationary in the distant past and in the distant future, in a certain sense that is much weaker than what is generally assumed in the CBC literature. We then recall how one can select canonical Poincar\'e subgroups $\Poi$ and $\Pof$ of $\B$ on $\scrip$ of such space-times using this asymptotic stationarity in the past (i.e. as one approaches $i^{o}$ along $\scrip$) and  future (i.e. as one approaches $i^{+}$ along $\scrip$); see Fig. 1. The past total angular momentum of the system $\Ji$ and the spin of the final black hole $\Sf$ refer to these two Poincar\'e subgroups, respectively. In the second part we present the expressions of $\Ji$ and $\Sf$. In the third, we discuss the non-triviality involved in comparing $\Ji$ and $\Sf$. The results are instructive for both mathematical relativists and gravitational wave  
theorists. 

\subsection{Poincar\'e subgroups in the distant past and distant future}
\label{s3.1}

We now wish to consider systems which do allow gravitational waves --so the Bondi news $N\, \equiv\, - \dot{\bar{\sigma}}^{\circ}$  on $\scrip$ is non-zero.  Therefore, we will impose the two conditions introduced in Section \ref{s2.3}  \emph{only in the limits $u\to \pm \infty$}. Throughout, we assume that if a field $F(u,\theta,\phi) = O(1/|u|^{\alpha})$\,\, --i.e., if  $|u|^{\alpha}F(u,\theta,\varphi)$ admits smooth limits $F_{\pm}(\theta,\varphi)$ as $u\to \pm\infty$-- \,\,then its $m$th\, $u$-derivative, $\partial_{u}^{m} F(u,\theta,\varphi)$  is $O(1/|u|^{m+\alpha})$. Then our conditions will be:
\begin{enumerate}[label=(\roman*)]
\item \emph{the Bondi news  $N \equiv -\dot{\bar\sigma}$ along $\scrip$ goes to zero as $u\to \pm \infty$ as $1/|u|^{1+\epsilon}$ for some $\epsilon >0$}; i.e.  $N$ is $O(1/|u|^{1+\epsilon})$, and
\item \emph{$\partial_{u} \1z \to 0$ in the past Bondi-frame as $u\to -\infty$, and in the future Bondi-frame as  $u\to \infty$.}
\end{enumerate}
In condition (ii), the future Bondi-frame is the one in which the future limit of the Bondi 3-momentum vanishes; thus it corresponds to the future rest frame of the system.  Similarly the past Bondi-frame corresponds to the past rest frame of the system.  Generically, the two are distinct. In CBCs, evolution in the future part of the coalescence is calculated using NR. Within the numerical accuracy, $\dot{\bar\sigma}^{\circ}$ goes to zero rapidly after the merger and the final state of the system is well described by a Kerr black hole for which we also have $\partial_{u}\, \1z =0$ (see, e.g. \cite{nblr}). Therefore conditions (i) and (ii) are readily satisfied in the distant future. The evolution of the binary in the distant past is calculated using PN methods where it is assumed that the system is stationary to the past of some time $t=-\tau$ (see, e.g., \cite{blanchet1}). Therefore our two conditions are trivially satisfied also in the distant past. Thus, space-times under consideration include \emph{in particular} those used to create CBC waveforms for detection and source characterization. For further discussion on motivation and implications of these two asymptotic conditions, see the companion paper \cite{adlk1}. Following  terminology used in that paper,  space-times satisfying conditions (i) and (ii) will be said to be \emph{past and future tame on $\scrip$}. In this Section we will work with this class of space-times. Now, while condition (i) is essential to ensure finiteness of the flux of energy momentum and angular momentum across $\scrip$, condition (ii) is not as compelling.  Therefore, in Appendix \ref{a1} we will weaken it. We will find that the procedure used in section \ref{s3.3} still goes through under these weaker conditions  but one cannot draw the conclusions we arrive at in section \ref{s4}.\vskip0.15cm

Considerations of Section \ref{s2.3} do not hold on all of $\scrip$ in space-times that are past and future tame. But they do hold in the limits  $u\to \pm\infty$. More precisely, various fields have the following asymptotic behavior in any Bondi-frame:
\begin{enumerate}
\item The waveform $2\sigmaz = \lim_{r\to\infty}\, r \big(h_{+} + ih_{\times}\big)$ has the asymptotic form:
\be \label{assum1}  \sigmaz (u,\theta,\varphi) = \sigma_{\pm} (\theta,\varphi) + |u|^{-\epsilon} \, \sigma^{(1)}_{\pm} (\theta,\varphi) + O (|u|^{-\epsilon-1} ) \quad {\rm as}\,\, u\to \pm\infty \ee
\vskip0.15cm
\noindent so that $\sigma_{\pm} (\theta,\varphi)$ are the limits of $\sigmaz (u,\theta,\varphi)$, and $\mp\,\epsilon\, \sigma^{(1)}_{\pm} (\theta,\varphi)$ are the limits of $ |u|^{1+\epsilon}\, \bar{N}(u,\theta,\varphi)$ as $u \to \pm \infty$. These limits depend on the choice of the affine parameter $u$, i.e. on the choice of cross-section $u=0$ in the given Bondi-frame. In particular, if we make a (finite)  supertranslation 
\be \label{sigmatrans} u\to \t{u} = u + \s(\theta,\varphi), \quad  {\rm then} \quad \t\sigma_{\pm} (\theta,\varphi)= \sigma_{\pm}(\theta,\varphi) + \eth^{2} \s(\theta,\varphi). \ee
Note that the difference $[\sigmaz]_{i^{+}}^{\iz}$ is invariant under  supertranslations.
\item Bianchi identities and Einstein's equations imply that the Newman-Penrose component $\2z$ of the Weyl tensor (whose real part determines the Bondi 4-momentum in  the limits $u\to \pm\infty$) has the asymptotic form:
\be \label{psi2}  \2z (u,\theta,\varphi) = \psi_{\pm}(\theta,\varphi) + |u|^{-\epsilon}\, \psi_{\pm}^{(1)} (\theta,\varphi) + O (|u|^{-\epsilon-1} ) \quad {\rm as}\,\, u\to \pm\infty \, ,\ee
where the limiting values $\psi_{\pm}(\theta,\varphi)$ and $\psi_{\pm}^{(1)} (\theta,\varphi)$ depend on the choice of the Bondi-frame, but $\psi_{\pm}(\theta,\varphi)$ is real in all Bondi-frames, $\psi_{-}$ is spherically symmetric in the past Bondi frame and $\psi_{+}$ in the future Bondi frame.
\item The  Newman-Penrose component $\1z$ of the Weyl tensor (which (together with $\Re \2z$) determines the BMS angular momentum as $u \to \pm\infty$) has the asymptotic form:
\be \label{psi1}  \1z (u,\theta,\varphi) = \chi_{\pm} (\theta,\varphi) + |u| \, \eth\psi_{\pm} (\theta,\varphi) + O (|u|^{-\epsilon} ) \quad {\rm as}\,\, u\to \pm\infty \, ,\ee
where the limiting value $\chi_{\pm}$ depends on the choice of the Bondi-frame as well as of the affine parameter $u$.
\end{enumerate}

Let us first consider the past limit, $u\to -\infty$.  Since  $\Im \2z = 2\sqrt{2} \, \Im \big(\eth^{2} \b{\sigma}^{\circ} + \sigmaz  \dot{\bar\sigma}^{\circ} \big)$ everywhere on $\scrip$, and in the limit   ${u\to -\infty} $, it follows \cite{adlk1} that \, $\Im \2z $ and $\dot\sigmaz(u,\theta,\phi)$  vanish in any Bondi-frame, condition (i) implies that the \emph{limiting value} $\sigma_{-}(\theta,\varphi)$ must satisfy  
\be \label{electic2}   \Im\, \eth^{2} {\bar\sigma}_{-} =0, \qquad {\hbox{\rm whose general solution is}} \quad \sigma_{-} = \eth^{2} {\s}_{-}    \ee
for some function $\s_{-}(\theta,\varphi)$. Therefore, given any Bondi-frame $(\qo_{ab}, \no^{a})$ we can use the transformation property (\ref{sigmatrans})  to choose an affine parameter $\t{u}$ such that the limiting value $\t{\sigma}_{-}(\theta,\varphi)$ on the cuts $\t{u} = const$ vanishes as $\t{u}\to -\infty$. Furthermore, since $\eth^{2} \s_{-} =0$ if and only if $\s_{-}(\theta,\varphi)\no^{a}$ is a BMS translation, there is precisely a 4-parameter family of cross-sections $\t{u} = const$ with the property $\lim_{\t{u}\to -\infty} \t{\sigma}_{-} =0$. We will refer to these cross-sections as \emph{cuts that become asymptotically `good' in the past}. Note that this 4-parameter family is uniquely chosen in any space-time that is past tame on $\scrip$; \emph{the family obtained starting from one Bondi-frame is the same as that obtained starting from another.} Since the family is preserved by translations but no other supertranslations, the subgroup of the BMS group $\B$ that preserves it is precisely a Poincar\'e group.  We will denote by $\Poi$. Note that for finite value of $\t{u}_{o}$, the cross-sections $\t{u} = \t{u}_{o}$ do carry shear because the Bondi news at $\scrip$ is non-zero. Shear $\t{\sigma}^{\circ} (\t{u},\theta,\varphi)$ of these cross-sections vanish only in the limit $\t{u}\to -\infty$. (See Fig. 1.)

To summarize, in space-times that are past tame, we can select a preferred Poincar\'e subgroup $\Poi$ of $\B$ by constructing the 4-parameter family of cross-sections $C^{-}$ of $\scrip$ on which the waveform $h_{+} + i h_{\times} = 2\sigmaz(u,\theta,\varphi)$  vanishes as $u \to -\infty$.  We can repeat the procedure for $u\to \infty$ and construct a 4-parameter family of cross-sections $C^{+}$, representing the \emph{cuts that become asymptotically `good' in the future}. This family is also left invariant by a Poincar\'e  subgroup $\Pof$. When would the two Poincar\'e subgroups $\Poi$ and $\Pof$ be the same? They would be the same if and only if the family of  cuts that becomes shear-free in the past also becomes shear-free in the asymptotic future. From the procedure outlined above, it is clear that for this to happen we would need $[\sigmaz]_{i^{+}}^{\iz} =0$  so that a \emph{single} supertranslation $f(\theta,\varphi)$ can transform away both $\sigmaz_{-}$ and $\sigmaz_{+}$. That is, the \emph{gravitational memory} must vanish:%
\footnote{In the mathematical GR literature, this is sometimes referred to as the \emph{`total'} gravitational memory --the sum of the `ordinary' and `null' contributions \cite{memory1}-- while in the quantum gravity literature \cite{as-book,acl}, it is called \emph{soft charge}.} 
\be \label{nomemory} \lim_{u\to\infty} \sigmaz(u,\theta,\varphi) \, -\,  \lim_{u\to-\infty} \sigmaz(u,\theta,\varphi) = 0\, .\ee
The left side is independent of the choice of the Bondi-frame.  As emphasized already in  the early literature on gravitational waves, Eq. (\ref{nomemory}) is a very stringent condition and will not be generically satisfied. Therefore, {even for the class of space-times that are  past and future tame, generically the Poincar\'e groups $\Poi$ and $\Pof$ are distinct.} The initial total angular momentum  $\Ji$ in the past refers to $\Poi$ and the final spin $\Sf$ refers to $\Pof$. As has been emphasized in the mathematical GR literature over the years \cite{np2,jw-rev,aa-rad,aa-bib}, this situation is qualitatively different from that in special relativity.  Finally, the fact that $\Poi$ and $\Pof$ are generically different is unrelated to the fact that the asymptotic rest frames in the past and future are also generically different: Generically the past and future Poincar\'e groups are distinct even in absence of a black hole kick \cite{kicks1,kicks2}.\\
  
 \emph{Remark:} As noted in Remark 2 at the end of Section \ref{s2}, an invariant characterization of the Poincar\'e subgroups $\P$ of $\B$ is provided by `classical vacua' --connections $[D_{o}]$ intrinsically defined on $\scrip$ for which curvature is trivial.  Consider asymptotically Minkowski space-times in which curvature of the connection $[D]$ induced on $\scrip$ is non-trivial --allowing for generic radiation-- but becomes trivial in the asymptotic past and future of $\scrip$:$[D] \to [D_{o}(\iz)]$ as $u\to -\infty$ and $[D] \to [D_{o}(i^{+})]$ as $u\to \infty$.  Then one can again select canonical Poincar\'e subgroups $\Poi$ and $\Pof$,\, but generically the two are distinct because $ [D_{o}(\iz) ] \not=  [D_{o}(i^{+})] $\,\,\cite{aa-rad,aa-bib}. Since triviality of curvature requires only $N$ and $\Im \2z$ to be zero, the asymptotic conditions needed to extract $\Poi$ and $\Pof$ from $\B$ are weaker than those satisfied by space-times that are past and future tame on $\scrip$: One can replace condition (ii) that features $\1z$ with just one it its consequences:  $\Im \2z$ goes to zero asymptotically.  Angular momentum can also be defined in this more general setting. However,  {as discussed in Appendix \ref{a1}, the definition then requires additional care in how the limits are taken.} 
 
\subsection{Angular momentum in the distant past and distant future}
\label{s3.2}

With the two Poincar\'e groups $\Poi$ and $\Pof$ at hand we now have the first necessary ingredient to define the two angular momentum vectors $\Ji$ and $\Sf$. Let us begin with  the past limit $u\to -\infty$ and spell out the step by step procedure to define the total angular momentum vector $\Ji$ in the distant past.

Recall first that the angular momentum vector refers to an ${\rm SO(3)}$ subgroup of the Poincar\'e group,  selected by the rest frame of the system. Now,  $N(u,\theta,\varphi) \to 0$ as $u\to -\infty$ because of condition (i), the past limit  $\momi$ of the Bondi 4-momentum is well-defined and given, from \eqref{mom}, by:
\be \label{mom2} \momi   := - \f{1}{4\pi G} \, \lim_{u_{o}\to -\infty} \oint_{u=u_{o}}\!\d2vo\, \alpha(\theta,\varphi)\,\, \Re\2z \, , \ee
where $\alpha(\theta,\varphi)\no^{a}$ are the BMS-translations.   $\momi$ equals the ADM 4-momentum of the space-time \cite{aaam1}. 
To locate the past rest frame of the system,  one can calculate $\momi$ in any Bondi-frame, and then use the 3-momentum to perform a boost to arrive at the desired Bondi frame $(\qo_{ab}, \no^{a})$ in which the limiting 3-momentum vanishes. The angular momentum $\Ji$ refers to the ${\rm SO(3)}$ subgroups selected by this rest frame. More precisely, from the 4-parameter family of cuts of $\scrip$ that become asymptotically good in the distant past, one selects a sub-family $u=const$ which is preserved by the flow generated by the BMS time translation $\no^{a}$, and finds the ${\rm SO(3)}$ generators 
\be \label{rot-gen} R^{\,a}{}_{(k)} =  \mathring{\epsilon}^{ab}\, \partial_{b} \beta_{(k)}, \ee 
where $(k) = 1,2,3$,\, $\mathring\epsilon^{ab}$ is the area-form on the $u=const$ cross-sections and $\beta_{(k)} = (\sin\theta \cos\varphi, \, \sin\theta\sin\varphi,\, \cos\theta)$ in terms of the spherical coordinates of the spherical unit 2-sphere metric $\qo_{ab}$ (see the form of general BMS generators given in Eq. (\ref{generators})). Then, the explicit expression of $ \Ji$ is given as a limit of integrals on the $u=const$ cross-sections  \cite{jw-rev,td,ak-thesis,aaak}
\be \label{J2}  \Ji^{\,(k)}= - \f{1}{4\pi G}\,\, \lim_{u_{o}\to -\infty} \oint_{u=u_{o}}\!\!\! \!\!\!\d2vo\,\, \Im\, \big[\1z(u,\theta,\varphi) \,\b{\eth}\beta_{(k)}(\theta,\varphi)\big]\, .\ee
where $\1z$ can be calculated from the asymptotic form of the PN metric in the distant past when the waveform vanishes.  Now, any two {of these} $u=const$ families are related by a spatial BMS translation, whence the  rotation generators $R^{\,a}{}_{(k)}$ also transform by picking up a spatial translation. However, since we are in the rest frame, the Bondi 3-momentum vanishes at $i^{\circ}$, whence  $\Ji^{\,(k)}$ does not change. Nonetheless, {since we have to make a choice in the actual calculation,}  we will use  the past center of mass frame: the $u=const$ family of cross-sections for which the past limit of the \emph{boost}-angular momentum vanishes. Note that the two steps --finding the rest and the center of mass frames adapted to $i^{\circ}$-- can always be carried out if {the Bondi news falls-off as $u\to -\infty$.}
Finally, since we do have a Poincar\'e group $\Poi$ adapted to $i^{\circ}$, we can use special relativistic considerations and say that the pair $(P_{\iz}^{(\alpha)}, \, J^{(k)}_{\iz})$ captures the full information about energy-momentum and angular momentum of the system in the distant past. 

We can repeat the procedure for $u\to \infty$ to obtain the future limit  $\Sf$ of the angular momentum vector representing the spin of the final black hole. As with the past limit, we have to proceed in two steps. First, we have to find the Bondi-frame $(\qo^{\prime}_{ab}, \no^{\prime\,a})$ in which the system is at rest in the asymptotic future, and in the second step a 1-parameter family of cross-sections $u^{\prime} = const$ whose shear vanishes in the distant future, where $u^{\prime}$ satisfies $\no^{\prime\,a} \partial_{a} u^{\prime} =1$. %
\footnote{Since it is not necessary, we will not take the extra step to require that this family should represent the `center of mass foliation' in the future.}
 The spin $\Sf$ of the final black hole is the angular momentum vector that refers to the ${\rm SO(3)}$ subgroups of $\Pof$, selected by $(\qo^{\,\prime}_{ab}, \no^{\,\prime\,a})$, given by the obvious modifications of (\ref{J2}). Since the NR simulations show that the space-time geometry quickly approaches that of a Kerr solution in distant future, all the ingredients needed in this calculation can be extracted from the simulations.

As their definitions make it clear, $\Ji$ and $\Sf$ refers to specific ${\rm SO(3)}$ subgroups of $\Poi$ and $\Pof$, respectively. Generically the two Poincar\'e groups are distinct and therefore do not share \emph{any} ${\rm SO(3)}$ subgroups. Thus generically  $\Ji$ and $\Sf$ refer to distinct BMS generators. Therefore, if one were to succumb to the temptation of subtracting $\Sf$ from $\Ji$ to calculate the angular momentum vector radiated away, one would be `subtracting apples from oranges' and the result would be conceptually meaningless. What then is the relation between $\Ji, \, \Sf$ and the `angular momentum carried by gravitational waves'? We analyze this issue in the next Subsection.

\subsection{Relation between the initial and final angular momentum vectors}
\label{s3.3}

Recall that $\Ji$ refers to ${\rm SO(3)}$ subgroups of $\Poi$ selected by the past rest frame $(\qo_{ab}, \, \no^{a})$, while $\Sf$ refers to ${\rm SO(3)}$ subgroups of $\Pof$ selected by the future rest frame $(\qo_{ab}^{\,\prime}, \, \no^{\,\prime\,a})$. Therefore the relation between them involves the (total) gravitational memory  $[\sigmaz]^{\iz}_{i^{+}}$\, --the difference between asymptotic shears at $\iz$ and $i^{+}$-- \,that determines the supertranslation relating $\Poi$ and $\Pof$, as well as to the black hole kick that characterizes the change in the two asymptotic rest frames. Thus, we are led to consider 4 cases: (i)  $[\sigmaz]^{\iz}_{i^{+}}=0$ and zero kick; \,  (ii)  $[\sigmaz]^{\iz}_{i^{+}}=0$ and non-zero kick;\, (iii)  $[\sigmaz]^{\iz}_{i^{+}}\not=0$ and zero kick; \,and, (iv) $[\sigmaz]^{\iz}_{i^{+}} \not=0$ and non-zero kick.\, We will find that in the first three cases, the simple-minded procedure to calculate flux of angular momentum using $\Ji$ and $\Sf$ turns out to give the correct answer. For case (iii) this result is unexpected because one would have expected a supermomentum contribution to the flux, associated with the supertranslation relating $\Poi$ and $\Pof$. Vanishing of this supermomentum contribution is a consequence of the assumption that the system becomes asymptotic stationarity as $u\to \pm \infty$, introduced in section \ref{s3.1}. But  the first three cases are exceptional for CBC in that they involve extreme fine tuning of parameters characterizing the compact binary.  Case (iv) represents the generic situation. In this case, not only is the naive procedure conceptually incorrect but would also yield incorrect flux precisely because it ignores the supermomentum contribution.

 {Since it is the generic case that is of direct interest, logically it would be sufficient to just discuss the case (iv) and read off the consequences in the first three cases from the general result (\ref{finalreln1}). However,  the actual calculation in case (iv)  involves a number of intermediate steps that are easier to follow if one discusses the special cases first.  Therefore, we will adopt the pedagogical --rather than the most direct-- route and pass from special cases to the general one.}
 
 \subsubsection{ Cases (i) and (ii):  Zero gravitational memory } 
\label{s3.3.1}
If the gravitational memory vanishes, the past Poincar\'e group is the same as the future one, $\Poi = \Pof$,  and so the supertranslation ambiguity simply disappears. In case (i), the kick also vanishes, whence the future and past Bondi frames also coincide. Therefore the situation trivializes: Since $\Ji^{\,(k)}$ and  $\Sf^{\,(k)}$ refer to the \emph{same} ${\rm SO(3)}$ subgroup of the BMS group, it is meaningful to subtract the two to obtain the flux of angular momentum radiated across $\scrip$ which also refers to the same ${\rm SO(3)}$:
\be \label{reln1} \Sf^{\,(k)} - \Ji^{\,(k)} = \F_{(R_{(k)})}  \ee
where $\F_{(R_{(k)})}$ is the flux across $\scrip$ of the angular momentum associated with the BMS rotations $R^{\,a}_{(k)}$ that generate the ${\rm SO(3)}$ under consideration (see Eq. (\ref{rot-gen})). The flux $ \F_{(R_{(k)})}$ is given by \cite{aams} 
\be \label{flux3}  \F_{(R_{(k)})}  = -\, \f{1}{8\pi G} \int_{\scrip}\!\!\!\! \rmd u\,\, \d2vo\,\, \dot{\sigma}^{\circ\, ab} \big(\Lie_{R_{(k)}} \,\sigma^{\circ}_{ab} \big)\, .\ee

In the case (ii), we again have a preferred Poincar\'e subgroup: $\Poi = \Pof = \P$, say. However, this case is a bit more complicated technically because now the past and the future rest frames are different: $\Ji^{\,(k)}$ refers to the ${\rm SO(3)} $ subgroups associated with the past rest frame, while $\Sf^{\,(k)}$ refers to a ${\rm SO(3)}$ subgroups selected by the future rest frame. However,  since the two ${\rm SO(3)}$ subgroups belong to the same Poincar\'e group, conceptually the situation is the same as in special relativity: the future rotation generators can be taken to be linear combinations of past rotations and boosts. 

Recall that we are working in the (rest and the) center of mass frame adapted to $i^{\circ}$. Therefore, $u= const$ is the preferred foliation by cuts that become shear-free in the distant past, and for which the limiting  boost angular momentum $\Ji^{\star}$ vanishes. Let us suppose that the black hole kick --or the boost relating the past and future rest frames-- is in the $x$ (or first) direction with velocity $v$.%
\footnote {A boost in a general direction is discussed in the Remark at the end of this subsection.}
Then, in the Lie algebra of the Poincar\'e group $\P$, the future rotation generators  $R^{\,\prime\,a}_{(i)}$   are related to the past rotation generators $R^{a}_{(i)}$ and the past boost generators $K^{a}_{(i)}$ as follows:
\be \label{Rreln1} R^{\,\prime\,a}_{(1)} = R^{a}_{(1)};\quad
R^{\,\prime\,a}_{(2)} =  \gamma \,\big(R^{a}_{(2)} + v\, K^{a}_{(3)}\big); \quad
R^{\,\prime\,a}_{(3)} =  \gamma\, \big(R^{a}_{(3)} - v\, K^{a}_{(2)}\big)\, ,\quad
\ee
where the expression of the rotations $R^{a}_{(k)}$ is given by (\ref{rot-gen}) and of boosts $K^{a}_{(k)}$ by
\be \label{boost-gen}
K^{a}_{(k)} = \kappa_{(k)}(\theta,\varphi) n^{a} + \qo^{ab} D_{b} \kappa_{(k)}(\theta,\varphi)
\ee
with $\kappa_{(k)} = (\sin\theta\cos\phi,\sin\theta\sin\phi, \cos\theta)$. These 
 relations translate directly to angular momenta. Thus,  the spin vector $\Sf$ of the final black hole and the total angular momentum $\Ji$ of the binary in the distant past are related via:
\ba \Sf^{(1)} =  \vec{J}^{(1)}|_{i^{+}};\quad\quad
 \Sf^{(2)} &=&\gamma\,\big( \vec{J}^{(2)}|_{i^{+}}\, +\, v\, \vec{J}^{\star\,(3)}|_{i^{+}}\big); 
\nonumber\\
{\rm and} \quad \quad  \Sf^{(3)} &=& \gamma\,\big(\vec{J}^{(3)}|_{i^{+}}\, - \, v\, \vec{J}^{\star\,(2)}|_{i^{+}}\big);\quad
 \ea
where  $\vec{J}^{(k)}|_{i^{+}}$ and $\vec{J}^{\star\, (k)}|_{i^{+}}$ are evaluated at $i^{+}$ but refer to rotations and boosts defined in the past frame (while $\Sf^{(k)}$, of course,  refers to the rotations defined in the future rest frame). Now,
\be   \vec{J}^{\,(k)}|_{i^{+}} - \Ji^{\,(k)} = \F_{(R_{(k)})} \quad  {\rm and} \quad   \vec{J}^{\star\,(k)}|_{i^{+}} - \Ji^{\star\,(k)}  =  \F_{(K_{(k)})}\, ,\ee
 where $ \F_{(R_{(k)})}$ and $\F_{(K_{(k)})}$ denote the flux across $\scrip$ of angular momentum defined by the {generators $R^{a}_{(k)}$ of rotations and $K^{a}_{(k)}$ of boosts (see Eqs. (\ref{rot-gen}) and (\ref{boost-gen})). } Their expressions \cite{aams} are given, respectively, by  Eq. (\ref{flux3}) and
 \be
  \label{flux4}  \F_{(K_{\,(k)} )} =   -\, \f{1}{8\pi G} \int_{\scrip}\!\!\!\! \rmd u\,\, \d2vo\,\, \dot{\sigma}^{\circ\, ab} \big(\Lie_{K_{(k)}} - \kappa_{(k)}(\theta,\varphi)  \big)\sigma^{\circ}_{ab}\, .\ee
Finally, recall that we chose the foliation $u=const$ such that the boost momentum vanishes in the distant past. Now, the flux associated with a linear combination of two BMS generators is just that linear combination of the two fluxes. Therefore, we have:  
\ba 
\Sf^{\,(1)} - \Ji^{\,(1)}    =  \F_{(R_{\,(1)})};\quad\quad  
\gamma^{-1}\,\Sf^{\,(2)}  -  \Ji^{\,(2)}   &=&  \F_{(R_{(2)})} + v\, \F_{(K_{(3)})} ; \nonumber\\
{\rm and}\quad  \gamma^{-1}\,\Sf^{\,(3)}  - \Ji^{\,(3)}   &=&\F_{(R_{(3)})}  - v\, \F_{(K_{(2)})}\,.
\ea
This is the  desired  relation between the initial total angular momentum $\Ji^{\,(k)}$,\, the final spin $\Sf^{\,(k)}$, and the fluxes of rotational and boost angular momenta across $\scrip$:
 For components along the boost direction it is the same as Eq. (\ref{reln1}) in case (i) above, but for components in directions orthogonal to the boost direction the relation is different. But this phenomenon reflects just the standard, special relativistic effect of mixing of rotations and boost generators under the change of the rest frame. \\

\emph{Remark:} We chose the boost in the first direction just for concreteness. If we choose it to be along an arbitrary unit vector $\h{v}^{\ul{i}} $ in the 3-dimensional space  $S$ of space-translations then the relation (\ref{Rreln1}) between the future rotations and the past rotations and boosts generalizes:
\be \h{v}^{\ul{i}}\,R^{\,\prime\, a}_{\ul{i}} =   \h{v}^{\ul{i}}\, R^{a}_{\ul{i}}; \quad\quad 
\h{o}^{ \ul {i} }\,  R^{\,\prime\, a}_{ \ul {i}} = \gamma\, \h{o}^{ \ul {i}}\,\big(R^{a}_{\ul {i}} \, \, - v\, \epsilon_{ \ul {i}  \ul {j} \ul {k} } K^{a \ul {j}}\, \h{v}^{ \ul {k}} \big)\ee
where $\h{o}^{\ul{k}}$ is a unit vector in $S$ orthogonal to $\h{v}^{\ul{k}}$; \,\,$\h{v}^{\ul{i}} R^{a}_{\ul{i}}$ is the rotation along the axis defined by the unit vector $\h{v}^{\ul{i}}$;  $\h{o}_{\ul {i}} R^{a\,  \ul {i}}$  the rotation along the axis $\h{o}^{\ul{i}}$; and  
$\h{o}^{\ul{i}}\, \epsilon_{ \ul {i}  \ul {j} \ul {k} } K^{a \ul {j}}\, \h{v}^{ \ul {k}} $ the boost in the spatial direction orthogonal to both $\h{o}^{\ul{i}}$ and $\h{v}^{\ul{i}}$.  As a result, the non-trivial balance equations for angular momentum in the direction $\h{o}^{\ul{i}}$  orthogonal to the boost direction becomes
\be \label{general-direction}  \gamma^{-1}\,\Sf^{\,(o)} - \Ji^{\,(o)}    = \F_{(R_{(o)})}  - v  \,  \F_{(K_{(o^{\prime})})}
 \ee
where $\h{o}^{\prime}_{\ul(i)} = \epsilon_{\ul{i} \ul{j} \ul{k}}\,\h{o}^{\ul{j}} \, \h{v}^{\ul{k}}$ is the unit vector orthogonal to both $\h{o}^{\ul{j}}$ and  $\h{v}^{\ul{k}}$. 

\subsubsection{Case (iii): Non-zero gravitational memory and zero kick}
\label{s3.3.2}
In this case gravitational waves do not carry away net 3-momentum, whence the initial and the final rest frames are the same. Therefore for the entire calculation we can restrict ourselves to the  fixed Bondi-frame $(\qo_{ab}, \no^{a})$ that represents this common rest frame. Let us choose, as before,  the foliation $u=const$ of $\scrip$, the leaves of which become asymptotically  
{shear-free} and represent the center of mass frame \emph{in the past}.  The initial total angular momentum $\Ji^{\,(k)}$ refers to the ${\rm SO(3)}$ subgroup selected by this family.  Since the gravitational memory is non-zero,  for this family we have  $\sigma_{-}(\theta,\varphi) =0$ but $\sigma_{+}(\theta,\varphi) \not=0$ (in Eq. (\ref{assum1})).  Now, the final spin $\Sf$ refers to the Poincar\'e group $\Pof$ that preserves the 4-parameter family of cross-sections that are shear-free \emph{in the future}. These are obtained by going to a new foliation $\t{u} = const$, where
\be \label{super}  \t{u}=u + \s(\theta,\varphi)\quad  {\rm and} \quad  \eth^{2} \s (\theta,\varphi) =  [\sigmaz]^{i^{+}}_{\iz}  \equiv  \sigma_{+}(\theta,\varphi). \ee
As we saw before, $s(\theta,\varphi)$ is unique up to addition of a BMS translation $\alpha(\theta,\phi)$ (since they constitute the kernel of $\eth^{2}$). By varying $s$ in this permissible class, we obtain a 4-parameter family of cuts that become good in the asymptotic future. The future spin $\Sf$ refers to ${\rm SO(3)}$ subgroups adapted to a 1-parameter sub-family, $\t{u} = const$,  satisfying $\no^{a} \partial_{a} \t{u} = 1$. There is a 3-parameter family of such foliations, related by spatial translations. To select the rotation generators explicitly,  one needs to choose a specific foliation. One can easily obtain it by exploiting the facts that:  (i)  we have a preferred Bondi-frame $(\qo_{ab},\, \no^{a})$; and, (ii) \emph{given a Bondi-frame}, there is a well-defined notion of a `pure' supertranslation, i.e., one that is orthogonal to all the BMS translations: $\s_{\ell \ge 2}(\theta,\varphi) = \sum_{\ell=2}^{\infty} \sum_{m=-\ell}^{\ell} \, \s_{\ell, m} Y_{\ell,m}(\theta,\varphi)$. Therefore, a specific foliation adapted to the distant future can be obtained by setting $\s(\theta,\varphi) = \s_{\rm{\ell \ge 2}}(\theta,\varphi)$ in (\ref{super}).  Since $\t{u}$ satisfies $\no^{a}\partial_{a} \t{u} =1$, the final spin $\Sf$ can be obtained by using the rotational vector fields $\t{R}^{\,a}_{(i)}$ adapted to this foliation. The difference between these future  ${\rm SO(3)}$ generators $\t{R}^{\,a}_{(k)}$ and the past ones $R^{a}_{(k)}$ is a \emph{pure} supertranslation, determined by the fact that while $R^{a}_{(k)}$ are tangential to the $u= const$ cross-sections,  $\t{R}^{\,a}_{(k)}$ are tangential to the  $\t{u} = const$ cross-sections:
\be \label{relation1}  \t{R}^{\,a}_{(k)} = R^{a}_{(k)}  +  {\s} _{(k)}\, \no^{a},\quad {\rm where} \quad 
  \s _{(k)} = - \Lie_{R_{(k)}}\, \s_{\rm \ell \ge 2},  \ee
\smallskip
since $R^{\,a}_{(k)}\partial_{a} u = 0$ implies that $\t{R}^{\,a}_{(k)} \partial_{a} \t{u} =  0$.
Note that  since $\s_{\rm \ell\ge2}(\theta,\varphi)\no^{a}$  is a pure supertranslation in our Bondi frame, so is  $ {\s}_{(k)} (\theta,\varphi) \no^{a}$. 

{Now, conceptually one cannot subtract $\Sf^{\,(k)}$ from $\Ji^{\,(i)}$ to obtain the flux of angular momentum across $\scrip$ (as in (\ref{reln1}))  because the two vectors now refer to two different ${\rm SO(3)}$ subgroups of the BMS group $\B$, and in general the two subgroups need not have \emph{any} generators in common! But it is meaningful to subtract the angular momentum $\vec{J}^{\,(k)}|_{i^{+}}$ from $\J^{\,(k)}|_{\iz} \equiv \Ji^{\,(k)}$ since both correspond to the \emph{same}  rotation generators, $R_{(k)}^{\,a}$ associated with the \emph{asymptotic past}.  Then, the balance law for the BMS angular momenta implies 
\be \label{Jflux1} \vec{J}^{\,(k)}\big{|}_{i^{+}}  - \Ji^{\,(k)} =  \F_{(R_{(k)})}\, , \ee
where $ \F_{(R_{(k)})}$ is the flux of angular momentum associated with the generators  $R^{\,a}_{(k)}$ of the ${\rm SO(3)}$ subgroup adapted to $i^{\circ}$.  Now,  since the spin $\Sf^{\,(k)}$ of the final black hole is associated with the generators  $\t{R}^{\,a}_{(k)}$,  Eq. (\ref{relation1}) implies
\be \label{S-J} \Sf^{\,(k)} = \vec{J}^{\,(k)}\big{|}_{i^{+}} + P_{(\s_{(k)})}\big{|}_{i^{+}} \, ,\ee
where $P_{(\s_{(k)})}$ is the `supermomentum'  
\footnote{\label{fn5} Note that $\s_{\rm \ell \ge 2}$ in Eq. (\ref{super}) is a \emph{finite} supertranslation with physical dimensions of length. Therefore  $f_{\rm \ell \ge 2}$ also has physical dimensions of length. In the expression $f\no^{a}$ of infinitesimal generators of supertranslations used in Section \ref{s2.2}, on the other hand, $f$ is dimensionless and therefore $P_{(f)}$ of Eq. (\ref{supmom}) has dimensions of 4-momentum. {$P_{(\s_{(k)})}$, on the other hand, has dimensions of angular momentum because of the difference in physical dimensions of $f$ used in (\ref{supmom}) and $\s_{(k)}$.}}
obtained by substituting $f$ with $s_{(k)}$ in Eq. (\ref{supmom}). But because $ \s_{(k)} \no^{a}$ is a \emph{pure} supertranslation a simplification arises as follows. Condition (ii) implies that both $\2z|_{\iz}$ and $\2z|_{i^{+}}$ are spherically symmetric in the Bondi-frame in which the system is at rest both in the distant past and distant future. This fact together with condition (i) imply that the pure supermomentum component of Eq. (\ref{supmom}) vanish at $\iz$ as well as at $i^{+}$. Consequently, the flux $\F_{(f)}$ of  `pure' supermomentum also vanishes. For details see \cite{adlk1}.

Therefore, from (\ref{Jflux1}) and (\ref{S-J}) we conclude 
\be  \label{naive}   \Sf^{(k)}  - \Ji^{\,(k)}  =   \F_{(R_{(k)})} \, . \ee
Thus, because of the special circumstance that the flux of supermomentum associated with any `pure' supertranslation vanishes, the naive subtraction of the spin of the final black hole from the initial total angular momentum of the system gives one the flux of angular momentum associated with the rotation sub-group in the asymptotic past (or asymptotic future).  In terms of special relativity, the subtraction on the left side of (\ref{naive}) is like taking the difference between, say, the $z$-component $\vec{J}\cdot \h{z}$ of angular momentum at early time and the combination $(\vec{J}\cdot \h{z} - \vec{P}\cdot \h{y})$ at late time in Newtonian gravity. In general the result is physically meaningless. However, if the $y$ component of the 3-momentum $\vec{P}$ happens to vanish at late time, this subtraction does provide the correct answer for change in the $z$-component of angular momentum from early to late times.  The same phenomenon occurs in CBC if there is  no kick.

\subsubsection{Case (iv):  Non-zero gravitational memory  and non-zero kick}
\label{s3.3.3}
Finally, let us consider the generic case. Because the gravitational memory is non-vanishing, the past Poincar\'e subgroup $\Poi$ is distinct from the future one, $\Pof$, and because the kick is non-vanishing, the past rest frame $(\qo_{ab},\, \no^{a})$ is also distinct from the future one, 
$(\qo^{\,\prime}_{ab}, \, \no^{\prime\,a})$. As  a consequence, the situation is now more complicated both conceptually and technically in that we have to combine the steps used in Section \ref{s3.3.1} and Section \ref{s3.3.2}.

At $i^{\circ}$ we can begin as in Section \ref{s3.3.2}. Let us use the past rest and center of mass frames to define the initial angular momentum $\Ji^{\, (k)}$. Thus we have a 1-parameter family of cross-sections $u= const$ that become asymptotically shear free at $i^{\circ}$ (and satisfy $\no^{a} \partial_{a} u =1$).  For this family, the asymptotic shear $\sigma^{\circ}(u,\theta,\varphi)$ in the distant future is non-zero. Therefore, we can go to the foliation given by  $\t{u}  = u + \s_{\ell \ge 2}(\theta, \varphi)$ to obtain a new family of cross-sections $\t{u} = const$ which are asymptotically shear-free in the distant future. As in Section \ref{s3.3.2}, \,$\t{u}$ satisfies $\no^{a}\partial_{a}\, \t{u} =1$. However because of the kick,  now it is $\no^{\prime\, a}$ that is adapted to the future rest frame as in the case (ii) of Section \ref{s3.3.1}. Therefore, from the 4-parameter family of cuts that become asymptotically good in the distant future, we need to select a 1-parameter sub-family $u^{\prime} = const$,\, satisfying $\no^{\prime\,a} \partial_{a} u^{\prime} =1$. This family can be obtained simply by performing a boost on the  $\t{u}= const$ cross-sections (that are already good cuts in the asymptotic future). The precise boost is the one that relates the past Bondi-frame to the future one; as in Section \ref{s3.3.1} it is determined by the kick velocity $v$ of the final black hole, and the unit vector $\h{v}^{a}$ that specifies the direction of the kick.

The relation between  rotation generators $R^{\prime\,a}_{(k)}$ that are used to define the spin $\S_{f}^{(k)}$ of the final black hole and the rotation generators $\t{R}^{\,a}_{(k)}$ and boosts $\t{K}^{a}_{(k)}$ adapted to the $\t{u}=const$ cross-sections is the same as in Section \ref{s3.3.1}. We just have to replace $R^{\,a}_{(k)}$  and $K^{a}_{(k)}$ in Eq. (\ref{Rreln1}) with their tilde versions because now $u^{\prime} = const$ cuts are obtained by performing the boost on the $\t{u}=const$  cuts rather than on the $u=const$ cuts:
\be \label{Rreln2} R^{\,\prime\,a}_{(1)} = \t{R}^{a}_{(1)};\quad
R^{\,\prime\,a}_{(2)} =  \gamma\, \big(\t{R}^{\,a}_{(2)} + v\, \t{K}^{a}_{(3)}\big); \quad
R^{\,\prime\,a}_{(3)} =  \gamma\, \big(\t{R}^{\,a}_{(3)} - v\, \t{K}^{a}_{(2)}\big).\quad
\ee
 In the next step we relate $\t{R}^{a}_{(k)}$  and $\t{K}^{a}_{(k)}$ to their versions without a tilde using the supertranslation that relates the $\t{u} = const$ cuts (that become shear-free in the limit $u\to \infty$) with the $u=const$ cuts (that become shear-free in the limit $u\to -\infty$).  The two sets of rotations are related by a supertranslation, just as one would expect:
\be \label{Rreln3} \t{R}^{a}_{(k)} = R^{a}_{(k)} +  \s_{(k)} \no^{a}, \qquad {\rm where} \,\, \s_{(k)} = - \Lie_{R_{(k)}} \, \s_{\rm \ell \ge 2}. \ee
(Note that the relation guarantees that $\t{R}^{a} \partial_{a} \t{u} =0$.) To obtain the relation between the two sets of boosts we use the form (\ref{generators}) of the BMS generators, which tells us that the two sets of boosts have the following form: $\t{K}^{a}_{(k)} = \t{u} \,\kappa_{(k)}\,  \no^{a} + \mathring{\tilde{q}}^{ab} \partial_{b} \kappa_{(k)}$, where again $\kappa_{(k)}
\equiv (\sin\theta\,\cos\varphi,\, \sin\theta\, \sin\varphi,\, \cos\theta)$ and $ \mathring{\tilde{q}}^{ab}$ is the 2-sphere metric on the $\t{u} = const$ cross-sections; \, and  ${K}^{a}_{(k)}$ have the same form but without the tilde.  A straightforward calculation shows that: 
\be \label{Breln}\t{K}^{a}_{(k)} = K^{a}_{(k)}  + [\kappa_{(k)}\,\s_{\rm \ell \ge 2}  +\s^{\star}_{(k)}] \, \no^{a}, \quad {\rm where}\,\,  \s^{\star}_{(k)} = - \Lie_{K_{(k)}} \, \s_{\rm \ell \ge 2}. \ee 
Together, Eqs. (\ref{Rreln2}) and (\ref{Breln}) enable us to express the rotations $R^{\,\prime\,a}_{(k)}$ adapted to $i^{+}$ in terms of  rotations $R^{a}_{(k)}$ and boosts $K^{a}_{(k)}$ adapted to $i^{\circ}$:  
\be
\begin{split} \label{Rreln3} R^{\,\prime\,a}_{(1)} = {R}^{\,a}_{(1)} + \s_{(1)}\no^{a};\quad
R^{\,\prime\,a}_{(2)} &=  \gamma\, \big({R}^{\,a}_{(2)} + v\, {K}^{a}_{(3)} + g_{(2)} \no^{a}\big); \\
{\rm and} \quad R^{\,\prime\,a}_{(3)} &=  \gamma\, \big({R}^{\,a}_{(3)} - v\, {K}^{a}_{(2)} + g_{(3)} \no^{a}
\big)\, ;\quad
\end{split}\ee
\be \label{g}  {\rm where}\quad g_{(2)} = \s_{(2)} +v \,\kappa_{(3)}\, \s_{\rm \ell \ge 2} \,  + v \s^{\star}_{(3)}, \quad {\rm and} \quad 
g_{(3)} = \s_{(3)} -v \,\kappa_{(2)}\,\s_{\rm \ell \ge 2}  - v \s^{\star}_{(2)}\, .\ee
With relations (\ref{Rreln3}) at hand, as in Sections \ref{s3.3.1} and \ref{s3.3.2} we can express the difference  $\Sf^{(k)} - \Ji^{(k)} $ in terms of {fluxes  $\F_{(R_{(k)})},\, \F_{(K_{(k)})}$ of angular momentum, boost momentum and fluxes  $\F_{(\s_{(1)})}$ and $\F_{(g_{(k)})}$ of supermomentum} across $\scrip$, all of which can be computed directly from the waveforms (see Eqs.  (\ref{flux2}), (\ref{flux3}) and (\ref{flux4})):
\be
\begin{split} \label{finalreln1} 
 \Sf^{\,(1)} -\Ji^{\,\,(1)}= \F_{(R_{\,(1)})} + \F_{(\s_{(1)})}; \quad 
 \gamma^{-1}\Sf^{\,(2)}  - \Ji^{\,\,(2)}  &= \F_{(R_{(2)})}  + v \F_{(K_{(3)})} + \F_{(g_{(2)})};
 \\
\gamma^{-1}\Sf^{\,(3)} - \Ji^{\,\,(3)}    &= \F_{(R_{(3)})}  - v \F_{(K_{(2)})}+ \F_{(g_{(3)})} .
\end{split}
\ee 
Presence of supermomentum fluxes on the right side implies that, in the generic case, one cannot simply subtract the final spin $\Sf^{(k)}$ from the initial total angular momentum $\Ji^{(k)}$ of the binary to obtain the angular momentum radiated across $\scrip$.  The difference also involves fluxes of \emph{supermomentum}, in addition to the expected boost angular-momentum contribution that arises already in special relativity. An examination of the expression (\ref{g}) of $g_{(k)}$ shows immediately that this supermomentum  contribution vanishes identically if the the total memory vanishes \emph{or} the kick vanishes, as it must from our results of Sections \ref{s3.3.1} and \ref{s3.3.2}. 

However as we remarked earlier, the binary has to be fine tuned for either of these possibilities to occur. In the generic case, the presence of the supermomentum flux on the right side of (\ref{finalreln1})  is the concrete manifestation of the supertranslation ambiguity in the definition of angular momentum at $\scrip$, that has been emphasized over the years in the mathematical GR literature \cite{jw,bb1,crp,ms-thesis,ms,jw-rev,aams,rgjw,aajw,rp2,shaw,ds,td,wz,ak-thesis,aaak}.  This is a key signature of the surprising enlargement of the 10 dimensional Poincar\'e group $\P$ to the infinite dimensional BMS group $\B$ that accompanies the presence of gravitational waves. But these conceptual considerations have remained qualitative. Now that we have a concrete expression (\ref{finalreln1}) in terms of memory and the  kick, we can estimate the observational significance of this ambiguity for the CBCs currently under investigation. We do so in the next Section.\\
\goodbreak

\emph{Remarks:}
\begin{enumerate}
\item Using the fact that $s_{\rm \ell \ge 2}(\theta,\varphi)$ is a \emph{pure} supertranslation (in the past Bondi frame), it follows that  in their spherical harmonic decomposition,  $\s_{(k)}$ have only $\ell \ge2$ parts, and  $g_{(k)}$ have no $Y_{0,0}$ part and their $Y_{1,m}$-part is proportional to $v$. This fact will play a key role in Section \ref{s4} in estimating the supermomentum contribution to Eq. (\ref{finalreln1}).

\item Recall that in the distant past $\2z$ is spherically symmetric in the past Bondi-frame $(\qo_{ab}, \no^{a})$ that most of our analysis refers to. Therefore it follows that 
\be
\begin{split} P_{(g_{(k)})}|_{i^{\circ}} &= - \f{1}{4\pi G}  \lim_{u_{o}\to -\infty} \, \oint_{u=u_{o}} \!\d2vo\, g_{(k)}(\theta,\varphi)\,\, \Re\2z \,   = 0;\quad {\rm and},\\
P_{(f_{(k)})}|_{i^{\circ}} &= - \f{1}{4\pi G}  \lim_{u_{o}\to -\infty} \, \oint_{u=u_{o}} \!\d2vo\, f_{(k)}(\theta,\varphi)\,\, \Re\2z \,   = 0\, .\end{split}\ee
Consequently, in (\ref{finalreln1}) we can replace the fluxes of supermomenta with supermomenta evaluated at  $i^{+}$ to obtain:
\be
\begin{split} \label{finalreln2}
\Sf^{(1)}  - \Ji^{(1)}   = \F_{(R_{(1)})} + P_{(\s_{(1)})}|_{i^{+}}; \quad 
\gamma^{-1}\,\Sf^{(2)}  - \Ji^{(2)}   &=  \F_{(R_{(2)})}  + v \F_{(K_{(3)})} + P_{(g_{(2)})}|_{i^{+}}; \\
 \gamma^{-1}\, \Sf^{(3)} - \Ji^{(3)}  &= \F_{(R_{(3)})}  - v \F_{(K_{(2)})} +P_{(g_{(3)})}|_{i^{+}}.  
\end{split}
\ee  
\end{enumerate}

\section{Discussion}
\label{s4}

A quintessential feature of gravitational waves in full, non-linear GR (with zero cosmological constant) is that although  space-times representing isolated gravitating systems are asymptotically Minkowskian, the asymptotic symmetry group is enlarged from the isometry group $\P$ of Minkowski space to the infinite dimensional BMS group $\B$.  As a result, given a cross section $C$ of $\scrip$ representing a retarded time instant, and a generator $\xi^{a}$ of the BMS group, we have a  2-sphere integral $P_{(\xi)}[C]$ representing the  $\xi$-component of the  BMS momentum at that  retarded time instant. Similarly, we have a 3-surface integral $\F_{(\xi)}$ over $\scrip$, representing the flux of this BMS momentum carried by gravitational (and other) waves across $\scrip$. Einstein's equations and Bianchi identities imply that these integrals satisfy the expected balance law: $P_{(\xi)}|_{i^{\circ}} - P_{(\xi)}|_{i^{+}} = \F_{(\xi)}$.  Given a  BMS supertranslation $\xi^{a} = f(\theta,\varphi) \no^{a}$,  the 2-surface integral provides a supermomentum component, $P_{(f)}[C]$. In general, two BMS symmetries $R^{a}_{(k)}$ and $R^{\prime\, a}_{(k)}$ that generate rotations around the $(k)$th axis  
differ by a supertranslation, say $f\no^{a}$. Consequently, the corresponding angular momenta are related  by supermomentum: $\vec{J}_{(R^{\prime}_{(k)})}[C]  - \vec{J}_{(R_{(k)})}[C] = P_{(f)}[C]$.  In the mathematical GR literature, one works with $P_{(\xi)}[C]$ and $\F_{(\xi)}$ associated with every BMS generator $\xi^{a}$, and regards the `supertranslation ambiguity' in the notion of angular momentum simply as an inevitable consequence of the presence of gravitational waves, that one just has to live with. This is of course a fully consistent viewpoint. However, it does not help one \emph{relate fluxes of physical quantities across $\scrip$ with observables associated with sources,} an issue of central importance to the community investigating CBCs.  

In the Post-Newtonian and effective one body approximations, for example, one associates with sources energy-momentum 4-vectors $P_{a}$ and angular momentum 3-vectors $\vec{J}^{a}$, and uses balance laws to arrive at equations of motion of the binary, and waveforms it produces at $\scrip$.  But in these treatments, there is no mention of the supermomentum, or of the supertranslation ambiguity while discussing  angular momentum. 

How can one reconcile this with the BMS group and the notion of BMS momenta at $\scrip$? We addressed this issue in detail in section \ref{s3.1} and showed that the boundary conditions assumed by the CBC community  in the distant past naturally enable one to select a preferred Poincar\'e subgroup $\Poi$ of the BMS group $\B$ using constructions available in the literature \cite{rp2,aa-rad}. Therefore, one can restrict oneself only to those BMS generators $\xi^{a}$ that belong to this Poincar\'e group and speak of the associated Poincar\'e momentum --i.e. $P_{a}$ and $\vec{J}^{a}$--  as one does for special relativistic systems (see section \ref{s2.1}).  These notions are appropriate for the investigation of the motion of sources referred to above. However, as is well-known in the mathematical GR community, there is a catch. The boundary conditions that hold at $i^{+}$ in CBCs also select a Poincar\'e subgroup $\Pof$ of  $\B$ and generically it is distinct from $\Poi$, related to it  by a \emph{supertranslation}. Therefore, if one starts with the total angular momentum of the binary $\Ji^{(k)}$  in the distant past and uses balance laws associated only with the BMS generators in $\Poi$, \emph{one would not obtain the angular momentum (i.e. spin) $\Sf^{(k)}$ of the final black hole:} The two would be related by a component of supermomentum that is not captured in the flux of the Poincar\'e momenta considered: to relate $\Sf^{(k)}$ with $\Ji^{(k)}$, we are forced to analyze supermomenta. 

We carried out this analysis in detail in section \ref{s3.3}. The final expression (\ref{finalreln2}) provides an explicit formula for the supermomentum terms that must be included to calculate $\Sf$ from $\Ji$. For definiteness, we assumed that the kick is in the $x$ direction (which corresponds to $(k) =1$ in our notation) and found:
\be
\begin{split} \label{finalreln3}
\Sf^{(1)}  =  \Ji^{(1)}   +\F_{(R_{(1)})} + P_{(\s_{(1)})}|_{i^{+}}; \quad 
\gamma^{-1}\,\Sf^{(2)}  &=  \Ji^{(2)}  +\F_{(R_{(2)})}  + v \F_{(K_{(3)})} + P_{(g_{(2)})}|_{i^{+}};\\
\gamma^{-1}\,\Sf^{(3)}  &= \Ji^{(3)}  + \F_{(R_{(3)})}  - v \F_{(K_{(2)})} + P_{(g_{(3)})}|_{i^{+}}.  
\end{split}\ee 
{(For a boost in the general direction, see Eq. (\ref{general-direction}).)} The fluxes of angular momentum $\F_{(R_{(k)})}$  and boost angular momentum $\F_{(K_{(k)})}$ on the right side are expected already from special relativistic considerations. The supermomentum terms $P_{(\s_{(k)})}|_{i^{+}}$ and 
$P_{(g_{(k)})}|_{i^{+}}$, on the other hand, are the quintessential signatures of the enlargement of the Poincar\'e group $\P$ to the BMS group $\B$.%
\footnote{As explained in footnote \ref{fn5}, this supermomentum has dimensions of angular momentum. Its magnitude depends on both the (total) gravitational memory and the kick of the final black hole. In the generic case, both are non-zero.}
The supertranslations  
labeling the supermomenta are determined by the (finite) supertranslation $\s_{\rm \ell \ge2}$ relating the Poincar\'e groups $\Poi$ and $\Pof$ and the generators $R^{a}_{(k)}, \, K^{a}_{(k)}$ of rotations and boosts adapted to $i^{\circ}$:
\be
\begin{split}
 \s_{(k)}  &= - \Lie_{R_{(k)}} \s_{\rm \ell \ge2}; \qquad  \s^{\star}_{(k)}  = - \Lie_{K_{(k)}} \s_{\rm \ell \ge2};\\
 g_{(2)} &= \s_{(2)} +v \,\kappa_{(3)}\, \s_{\rm \ell \ge 2} \,  + v \s^{\star}_{(3)}, \quad {\rm and} \quad 
g_{(3)} = \s_{(3)} -v \,\kappa_{(2)}\,\s_{\rm \ell \ge 2}  - v \s^{\star}_{(2)}\, ,
\end{split}\ee
where the functions $\kappa_{(k)}(\theta,\varphi)$ defining the boosts  $K^{a}_{(k)}$ are given by $\kappa_{(k)} = (\sin\theta\cos\phi,$\, $\sin\theta\sin\phi,\, \cos\theta)$. (For explicit expressions for rotations and boosts, see  (\ref{rot-gen}) and (\ref{boost-gen}).)  

With this explicit expression at hand, one can estimate the magnitude of the supermomentum using the general expression (\ref{supmom})
\be P_{(f)}|_{i^{+}}   := - \f{1}{4\pi G} \,\lim_{u\to\infty} \oint_{u=u_{o}}\!\!\d2vo\,  f(\theta,\varphi)\, \Re \2z (\theta,\varphi)\, ,
\ee
and substituting $\s_{(k)}$ and $g_{(k)}$ for $f$. Now, the boundary conditions at $i^{+}$ of section \ref{s3.1} imply that, in the Bondi-frame adapted to $i^{+}$, \,$\2z|_{i^{+}}$ is spherically symmetric. {Hence by the well-known transformation property, in the Bondi-frame adapted to $i^{\circ}$ used  in the paper, we have:}
\be
\begin{split} \label{future2z} {\2z} \big{|}_{u=\infty} \, &=   -\, \f{GM_{i^{+}} (1-v^{2})^{\f{3}{2}} }{\big(1 - v \, \sin\theta\cos\varphi\big)^{3}}\\
&= -\,GM_{i^{+}}\,\Big (1 + (3\sin\theta\cos\varphi)\, v  - \big(\f{3}{2} - 6 \sin^{2}\theta \cos^{2}\varphi\big) \, v^{2} + \ldots \Big),\end{split}\ee
{(as spelled out in \cite{adlk1}). Here} in the second step we truncated the Taylor expansion  in $v$, ignoring terms $O(v^{3})$ because numerical simulations show that the kicks are typically of the order of a few hundred kms/s, i.e. with $v \sim 10^{-3}$  (although one can, of course continue the expansion to higher orders). Now in the spherical harmonic decomposition, $\s_{(k)}$ and $g_{(k)}$ have no $Y_{00}$ parts;\,  $\s_{(k)}$ has no $Y_{1,m}$ part; and $\sin\theta\cos\varphi$ is a linear combination of $Y_{1,1}$ and $Y_{1,-1}$. Therefore the expressions of supermomenta $P_{(\s_{(1)})}$, $P_{(g_{(2)})}$ and  $P_{(g_{(3)})}$ simplify considerably; in terms of the kick velocity, their leading terms go as $\sim v^{2}$. Furthermore, analytic considerations \cite{garfinkle2016simple} as well as detailed calculations using available waveforms \cite{adlkk} show that the memory term, and hence $\s_{\rm \ell \ge 2}$ is of the order $O(G E_{\rm rad})$ where $E_{\rm rad}$ is the total energy radiated in the form of gravitational waves. Therefore, we obtain
\be P_{(\s_{(k)})} \sim P_{(g_{(k)})}  = M_{i^{+}} \, O(G E_{\rm rad}) \, v^{2}\, +\, O(v^{3})\, . \ee 

What is the magnitude of the error involved if these supermomenta are ignored in Eq. (\ref{finalreln3})? To assess the importance of these terms,  let us consider the  (dimensionless) \emph{fractional contributions}  $\t{P}_{(\s_{(k)})}$ and $\t{P}_{(g_{(k)})}$ of supermomenta to the total angular momentum flux:
\be \t{P}_{(\s_{(k)})}  =  \f{P_{(\s_{(k)})}}{|\Ji- \Sf|},   \quad{\rm and} \quad   \t{P}_{(g_{(k)})}  = \f{P_{(g_{(k)})}}{|\Ji- \Sf|}\, .
\ee
Now,  in binary coalescences, the initial total angular momentum is much larger than the spin of the final black hole, whence  $|\Ji- \Sf| > |\Sf| \sim M_{i^{+}}^{2}$. 
Therefore, the fractional supermomentum contribution to the angular momentum flux is bounded by:
\be \label{Delta} \t{P}_{(\s_{(k)})}  <\,\,   O \Big(\f{G E_{\rm rad}}{M_{i^{+}}}\Big) \, v^{2}\,  \quad {\rm and}\quad  
\t{P}_{(g_{(k)})} <\,\, O \Big(\f{G E_{\rm rad}}{M_{i^{+}}}\Big) \, v^{2}\,  .\ee
%
Now, since the fractional energy radiated in the form of gravitational waves is less than 10\%, the right sides in (\ref{Delta}) are less than $\sim\,10^{-1} v^{2}$. As we already noted,  typically $v\sim 10^{-3}$ in CBCs  that are usually considered. The largest black hole kicks that have been seen numerically have velocities of $\sim 5000$km/s, or, $v \sim 2 \times 10^{-2}$ \cite{kicks1,kicks2}.  Even for these, the fractional supermomentum contributions  would only be  $O(10^{-5})$.  Thus, while ignoring the supermomentum terms in the calculation of the angular momentum flux is conceptually incorrect, the numerical error is small for the CBCs of current interest. \vskip0.15cm

To summarize,  the detailed calculations of Section \ref{s3} provide a resolution of the tension surrounding angular momentum at $\scrip$ that should be satisfying to both the mathematical GR and waveform communities. As emphasized by the mathematical GR community, the supertranslation ambiguity is indeed inevitable because the BMS group $\B$ does \emph{not} admit a preferred Poincar\'e group.  Generically the total angular momentum $\Ji$ of the initial binary and the spin $\Sf$ of the final black hole  refer to two \emph{different} Poincar\'e subgroups, $\Poi$ and $\Pof$, of $\B$. One could imagine working with $\Poi$ and using its rotation and boost generators $R^{a}_{(k)}$ and $K^{a}_{(k)}$ to define the initial $\Ji$ and fluxes  $\F_{(R_{(k)})}$ and $\F_{(K_{(k)})}$ across $\scrip$. But, one cannot obtain the spin $\Sf$ of the final black hole by subtracting these fluxes from $\Ji$ because $\Sf$ refers to the rotations $R^{\prime\,a}_{(k)}$ in the future Poincar\'e group $\Pof$. Therefore,  in the generic case when the gravitational memory does not vanish, $R^{\prime\,a}_{(k)}$ are related to $R^{a}_{(k)}$ and $K^{a}_{(k)}$ via a supertranslation. Therefore, to obtain $\Sf$, one also needs the corresponding supermomentum flux. However it turns out that under asymptotic conditions in the distant future and past that are generally used in the analysis of CBCs,  the fractional contribution of this supermomentum flux to $\Sf$ is only $O(10^{-7})$ (for kick velocities $v \sim 10^{-3}$ that are normally encountered). Therefore, the error one would make by ignoring the supermomentum flux in the above calculation of $\Sf$ is too small to be relevant to the current gravitational wave detectors or those that will be built in foreseeable future. Thus, while it is conceptually incorrect to restrict oneself to just one Poincar\'e subgroup $\Poi$ of $\B$,  in practice one can do so for angular momentum considerations.%
\footnote{At first it seems surprising that the size of the  supermomentum contribution is dictated primarily by the kick velocity, and not by the size of the gravitational memory. This is because of the boundary conditions that are normally used as $u\to \pm \infty$, discussed in Section \ref{s3.1}. The total gravitational memory $[\sigma]^{i^{\circ}}_{i^{+}}$ is a sum of two contributions  --a linear (or ordinary) term $[\2z]^{i^{\circ}}_{i^{+}}$ and a non-linear (or null) term $\int  |\dot {\sigma}^{\circ}|^{2}\, \rmd u$. The second is essentially bounded by the total energy radiated \cite{garfinkle2016simple,adlkk}. The first could in principle be large but the boundary conditions tie the corresponding supermomentum to the kick velocity and make it $O(v^{2})$.} 

Note, however,  that there is a caveat: the asymptotic conditions normally imposed in the far past and far future played an essential role in arriving at this conclusion.  As we point out in Appendix \ref{a1}, it is possible to weaken these conditions and still carry out the analysis of section \ref{s3.3} and arrive at Eq. (\ref{finalreln3}). However, under these weaker conditions, we can no longer conclude from (\ref{finalreln3}) that supermomentum contributions are $O(v^{2})$;  a priori the quantities $\t{P}_{(\s_{(k)})}$ and  $\t{P}_{(g_{(k)})}$ of Eq. (\ref{Delta}) could well be $O(1)$. Thus, if it should turn out that the asymptotic stationary conditions normally assumed  are too strong for some CBCs, one would have to revisit the issue of the significance of supertranslation ambiguities using (\ref{finalreln3}). \vskip0.15cm

 {Finally, there is  a broader question for the geometric analysis community: Does there exist a sufficiently large class of space-times admitting CBCs that are asymptotically flat in the sense of Penrose \cite{rp}, so that $\1z$ is well-defined on $\scrip$ for all finite values of $u$? If $\1z$ fails to exist, then we will not even be able to define angular momentum $\J^{a}[C]$ at cross-sections $C$ of $\scrip$! In the early days  there were concerns, based on approximation methods, as to whether the underlying assumptions are too strong to be satisfied by realistic isolated systems such as compact binaries (see, e.g., \cite{ww,lbtd}). By now 
 there is a consensus in the physics community that there is indeed a large class of CBCs with sufficient regularity at $\scrip$ to make the notion of angular momentum well-defined. In particular, the asymptotic form of the PN metric is completely consistent with the Bondi-Sachs-Penrose framework, as shown for instance by Theorem 4 in \cite{blanchet1}.  What is the status of rigorous mathematical results?  There are powerful theorems on non-linear stability of Minkowski space \cite{dcsk,lb,pced} showing that the degree of regularity at  $\scrip$ depends on the precise fall-off conditions assumed for the initial data. However, it is not clear which of these conditions are the most appropriate ones for CBCs. For, in the analysis of compact binaries we are interested in solutions in which the Bondi news vanishes on $\scrim$, so that all the radiation at $\scrip$ is created by the binary. But the only space-time with this property that is covered by the available results on the non-linear stability of Minkowski space is Minkowski space itself. Consequently, as of now, the non-linear stability results  do not provide direct guidance for the issue of regularity at $\scrip$ in generic CBCs. }

 \section*{Acknowledgments}
We would like to thank  B. Krishnan and B. Schultz  for discussions. This work was supported in part by the NSF grant PHY-1806356, grant UN2017-92945 from the Urania Stott Fund of Pittsburgh Foundation and the Eberly research funds of Penn State.

\begin{appendix}
\section{Weakening condition (ii) of Section \ref{s3.1}}
\label{a1}

Throughout our analysis of Section \ref{s3.3}, we assumed that space-times under consideration are past and future tame on $\scrip$, i.e., that they satisfy conditions (i) and (ii) of Section \ref{s3.1} in the limit to $\iz$ \emph{and} $i^{+}$ along $\scrip$. These conditions are weaker than the asymptotic stationarity normally assumed in the CBC community. Still,  in light of our surprising finding in Section \ref{s4}  that the supermomentum contribution in Eq. (\ref{finalreln3}) is negligible in practice, it is appropriate to re-examine these assumptions.  As we commented after the introduction of these conditions, (i) is compelling on physical grounds because it is essential to guarantee that the total flux of energy momentum and angular momentum across $\scrip$ is finite. What about condition (ii)? Justification for it is not as compelling. Can we then perhaps weaken it? The answer is in the affirmative. We can replace (ii)  on $\partial_{u} \1z$ by a weaker condition that is a consequence of (ii):
\begin{enumerate}[label=(\roman*)${}^{\prime}$]
\setcounter{enumi}{1}
\item $\Im \2z \to 0$\, as $u\to \pm \infty$.
\end{enumerate}
As mentioned in the Remark at the end of Section \ref{s3.1}, this condition is directly motivated by considerations of `classical vacua' $[D_{o}]$. But note that (ii)${}^{\prime}$ is genuinely weaker than (ii): Since it allows $\partial_{u} \1z$ to be non-zero in the limits $u\to \pm \infty$, now $\1z$ may diverge in these limits. Nonetheless, condition (ii)${}^{\prime}$ is necessary and sufficient to ensure that:
\begin{enumerate}[label=(\alph*)]
\item We can introduce cuts that become asymptotically shear-free as $u\to \pm\infty$, so that we can single out Poincar\'e groups $\Poi$ and $\Pof$ in the distant past and the future; and,
\item The angular momentum integral (\ref{J2}) is well-defined in the limits $u\to \pm \infty$.
\end{enumerate}
As discussed in the Remark mentioned above, conclusion (a) is directly implied by condition (ii)${}^{\prime}$. Conclusion (b) can be verified simply by integrating the equation that relates $\dot{\Psi}_{1}^{\circ}$ to $\eth \2z$, shear, and news and using the equation that relates $\Im \2z$ to shear and news, both of which are well-known consequences of the field equations and Bianchi identities.  Alternatively, one can establish it using the fact that $\vec{J}^{(i)}[C]$ is manifestly well defined for a finite cross-section $[C]$ since $\1z$ is well-defined there, and the fact that fluxes of angular momentum between $[C]$ and $u=\pm \infty$ are finite in virtue of the assumption (i) on the fall-off of news.

Thus, with these weakened condition,  $\Sf^{(i)}$ and $\Ji^{(i)}$ continue to be well-defined and their relation continues to be given by (\ref{finalreln3}).  However,  since now $\partial_{u} \1z \not\to 0$ as $u \to \pm \infty$ in general,  the limits $\psi_{\pm} (\theta,\varphi)$ of $\2z$ are not necessarily spherically symmetric in the rest frames at $i^{+}$ and $i^{\circ}$, respectively. Therefore,  in general 
\be
\lim_{u \to -\infty} \2z  \not=  GM_{i^{\circ}} \qquad {\rm and} \qquad  \lim_{u \to \infty} \2z  \not=  \f{GM_{i^{+}} }{\gamma^3(1- v \sin\theta\cos\varphi)^3} \ee
 in the past Bondi-frame. The angular dependence of the limits of $\2z$ is unrestricted. Consequently, (the ordinary or linear memory can be arbitrarily large and) the supermomenta in (\ref{finalreln3}) need not be $O(v^{2})$. Indeed, they need not vanish even in absence of a kick.  It may turn out that there are physically interesting CBCs that violate condition (ii) --and therefore also the much stronger asymptotic stationarity condition used in the CBC literature. Then the supertranslation ambiguity in the notion of angular momentum could be too large to be negligible even for the current generation of detectors. 
  
  Discussion of this Appendix brings out the subtle interplay between boundary conditions in the asymptotic past and future, and physical effects.

\end{appendix}

\bibliographystyle{apsrev4-1}
\bibliography{references}

\end{document}